\documentclass[trackchanges,twocolumn]{aastex701}

\newcommand{\mcwilliams}{
    McWilliams Center for Cosmology and Astrophysics,
    Department of Physics,
    Carnegie Mellon University,
    5000 Forbes Avenue, Pittsburgh, PA 15213, USA
}
\newcommand{\lmu}{
    University Observatory, 
    Faculty of Physics, 
    Ludwig-Maximilians-Universität München, 
    Scheinerstr. 1, 81679 Munich, Germany
}

\newcommand{\origins}{
    Excellence Cluster ORIGINS,
    Boltzmannstr. 2,
    85748 Garching, Germany
}

\usepackage{amsmath}
\usepackage{longtable}
\usepackage{caption}
\usepackage{placeins}
\usepackage{float}

\begin{document}

\title{Electromagnetic Follow-up of the Sub-Solar Mass Gravitational Wave Candidate S251112cm: Kilonova Constraints and a Coincident IIb Supernova}

\author[0000-0002-9364-5419]{Xander J. Hall}
\affiliation{\mcwilliams}
\email[show]{xhall@cmu.edu}

\author[0000-0002-2184-6430]{Tomas Ahumada}
\affiliation{Cerro Tololo Inter-American Observatory/NSF NOIRLab, Casilla 603, La Serena, Chile}
\email{tomas.ahumada@noirlab.edu}

\author[0009-0008-2754-1946]{Julius Gassert}
\affiliation{\mcwilliams}
\affiliation{\lmu}
\email{julius.gassert@campus.lmu.de}

\author[0000-0002-6011-0530]{Antonella Palmese}
\affiliation{\mcwilliams}
\email{apalmese@andrew.cmu.edu}

\author[0000-0002-4670-7509]{Brian D.~Metzger}\email{bdm2129@columbia.edu}
\affiliation{Department of Physics and Columbia Astrophysics Laboratory, Columbia University, New York, NY 10027, USA}
\affiliation{Center for Computational Astrophysics, Flatiron Institute, 162 5th Ave, New York, NY 10010, USA}

\author[0000-0002-5619-4938]{Mansi M. Kasliwal}\email{mansi@astro.caltech.edu}
\affiliation{Division of Physics, Mathematics and Astronomy, California Institute of Technology, Pasadena, CA 91125, USA}

\author[0000-0002-8255-5127]{Mattia Bulla}\email{mattia.bulla@unife.it}
\affiliation{Department of Physics and Earth Science, University of Ferrara, via Saragat 1, I-44122 Ferrara, Italy} \affiliation{INFN, Sezione di Ferrara, via Saragat 1, I-44122 Ferrara, Italy} \affiliation{INAF, Osservatorio Astronomico d’Abruzzo, via Mentore Maggini snc, 64100 Teramo, Italy}

\author[0000-0003-3270-7644]{Daniel Gruen}
\affiliation{\lmu}
\affiliation{\origins}
\email{daniel.gruen@lmu.de}

\author[0000-0003-2434-0387]{Robert Stein}\email{rdstein@umd.edu}
\affiliation{Department of Astronomy, University of Maryland, College Park, MD 20742, USA} \affiliation{Joint Space-Science Institute, University of Maryland, College Park, MD 20742, USA} \affiliation{Astrophysics Science Division, NASA Goddard Space Flight Center, MC 661, Greenbelt, MD 20771, USA}

\author[0000-0002-4223-103X]{Christoffer Fremling}\email{fremling@caltech.edu}
\affiliation{Caltech Optical Observatories, California Institute of Technology, Pasadena, CA 91125, USA} \affiliation{Division of Physics, Mathematics and Astronomy, California Institute of Technology, Pasadena, CA 91125, USA}

\author[0000-0003-3768-7515]{Shreya Anand}\email{sanand08@stanford.edu}
\altaffiliation{LSST-DA Catalyst Postdoctoral Fellow} \affiliation{Kavli Institute for Particle Astrophysics and Cosmology, Stanford University, 452 Lomita Mall, Stanford, CA 94305, USA} \affiliation{Department of Astronomy, University of California, Berkeley, CA 94720-3411, USA}

\author[0000-0002-8977-1498]{Igor Andreoni}
\affiliation{University of North Carolina at Chapel Hill, 120 E. Cameron Ave., Chapel Hill, NC 27514, USA}
\email{igor.andreoni@unc.edu}

\author[0009-0001-0574-2332]{Malte Busmann}\altaffiliation{Recipient of a Wübben Stiftung Wissenschaft Student Grant}
\affiliation{\lmu}
\affiliation{\origins}
\email{m.busmann@physik.lmu.de}

\author[0000-0002-1270-7666]{Tom\'as Cabrera}
\affiliation{\mcwilliams}
\email{tcabrera@andrew.cmu.edu}

\author[0009-0000-6729-9340]{Ryan Christinzio}
\affiliation{\mcwilliams}
\email{rpchrist@andrew.cmu.edu}

\author[0009-0006-7990-0547]{James Freeburn}
\affiliation{Sydney Institute for Astronomy, School of Physics, University of Sydney, Sydney, NSW 2006, Australia} \affiliation{ARC Centre of Excellence for Gravitational Wave Discovery (OzGrav), Hawthorn, Victoria, 3122, Australia}
\affiliation{Department of Physics and Astronomy, University of North Carolina at Chapel Hill, Chapel Hill, NC 27599-3255, USA}
\email{jfreebur@unc.edu}

\author[0000-0003-2362-0459]{Ignacio Maga\~na~Hernandez}
\affiliation{\mcwilliams}
\email{imhernan@andrew.cmu.edu}

\author[0000-0001-7201-1938]{Lei Hu}
\affiliation{\mcwilliams}
\email{leihu@andrew.cmu.edu}

\author[0000-0002-9700-0036]{Brendan O'Connor}
\altaffiliation{McWilliams Fellow}
\affiliation{\mcwilliams}
\email{boconno2@andrew.cmu.edu}  

\author[0000-0002-9092-0593]{Ji-an Jiang}
\affiliation{Department of Astronomy, University of Science and Technology of China, Hefei 230026, People’s Republic of China}
\affiliation{School of Astronomy and Space Sciences, University of Science and Technology of China, Hefei 230026, People’s Republic of China}
\email{jian.jiang@ustc.edu.cn}

\author[0000-0002-2242-1514]{Zhengyan Liu}
\affiliation{Department of Astronomy, University of Science and Technology of China, Hefei 230026, People’s Republic of China}
\affiliation{School of Astronomy and Space Sciences, University of Science and Technology of China, Hefei 230026, People’s Republic of China}
\email{ustclzy@mail.ustc.edu.cn}

\author[0000-0002-1330-2329]{Wen Zhao}
\affiliation{Department of Astronomy, University of Science and Technology of China, Hefei 230026, People’s Republic of China}
\affiliation{School of Astronomy and Space Sciences, University of Science and Technology of China, Hefei 230026, People’s Republic of China}
\email{wzhao7@ustc.edu.cn}

\author[0000-0001-8018-5348]{Eric C. Bellm}
\email{ecbellm@uw.edu}
\affiliation{DIRAC Institute, Department of Astronomy, University of Washington, 3910 15th Avenue NE, Seattle, WA 98195, USA}

\author[0000-0002-6877-7655]{David Cook}
\email{dcook@ipac.caltech.edu}
\affiliation{IPAC, California Institute of Technology, 1200 E. California Blvd, Pasadena, CA 91125, USA}

\author[0000-0002-8262-2924]{Michael W. Coughlin}
\email{cough052@umn.edu}
\affiliation{School of Physics and Astronomy, University of Minnesota, Minneapolis, Minnesota 55455, USA}

\author[0000-0002-5884-7867]{Richard Dekany}
\affiliation{Caltech Optical Observatories, California Institute of Technology, Pasadena, CA  91125, USA}
\email{rgd@astro.caltech.edu}

\author[0000-0002-3168-0139]{Matthew Graham}
\affiliation{Cahill Center for Astronomy and Astrophysics, California Institute of Technology, Pasadena, CA 91125, USA}
\email{mjg@caltech.edu}

\author[0000-0003-2451-5482]{Russ R. Laher}
\affiliation{IPAC, California Institute of Technology, 1200 E. California Blvd, Pasadena, CA 91125, USA}
\email{laher@ipac.caltech.edu}

%% Use the \collaboration command to identify collaborations. This command
%% takes an optional argument that is either a number or the word "all"
%% which tells the compiler how many of the authors above the command to
%% show. For example "\collaboration[all]{(DELVE Collaboration)}" wil include
%% all the authors above this command.
%%
%% Mark off the abstract in the ``abstract'' environment. 
\begin{abstract}

On November 12th, 2025 the LIGO--Virgo--KAGRA (LVK) collaboration reported gravitational waves (GWs) from a compact object merger candidate (S251112cm) with at least one sub-solar mass component. Using the Dark Energy Camera (DECam), the Fraunhofer Telescope at Wendelstein Observatory (FTW), and the Zwicky Transient Facility (ZTF), we surveyed $56\%$ of the GW localization region beginning $2.4$~hours after the GW alert. We find no kilonova (KN) counterpart, and use radiative-transfer models to rule out $42\%$ (ZTF), $68\%$ (DECam), and $92\%$ (FTW) of the KN models as possible emission from this GW candidate. Within the recently proposed disk-fragmentation (``superkilonova'') model for generating sub-solar mass neutron star mergers from stellar core-collapse, the delay between the supernova explosion time and the GW merger time is estimated to be less than a few days. Searching this time window prior to the GW event, we identify and spectroscopically classify a IIb supernova (SN~2025adtq), with a spatial association odds ratio of $\log_{10}\mathcal{I} \approx 4.8$, a chance coincidence probability of ${\sim}2$--$9\%$, and an estimated explosion time ${\sim}2$ days prior to S251112cm. SN~2025adtq is the second Type~IIb supernova found in spatial and temporal coincidence with a sub-solar mass GW candidate, following the previously reported S250818k/SN~2025ulz association; jointly, we measure an odds ratio that favors the association hypothesis over the null, however, when conditioned on finding a coincident supernova by chance, the odds ratio disfavors association. Together, these results provide suggestive but inconclusive evidence for the superkilonova formation channel.

%the two pairs yield odds ratios of $\Lambda_{\mathrm{uncond}} = 4.3$ and $\Lambda_{\mathrm{cond}} = 0.041$; while

%Looking ahead and using S251112cm and SN 2025adtq as benchmarks, we determine that at least $2$ ($5$) associations between sub-solar mass mergers and IIb SN are required to reject the null hypothesis at the $3\sigma$ ($5\sigma$) level. SN~2025adtq is the second Type~IIb supernova found in spatial and temporal coincidence with a sub-solar mass GW candidate, lending to growing but still inconclusive support to the superkilonova formation channel.

\end{abstract}

%% Keywords should appear after the \end{abstract} command. 
%% The AAS Journals now uses Unified Astronomy Thesaurus (UAT) concepts:
%% https://astrothesaurus.org
%% You will be asked to selected these concepts during the submission process
%% but this old "keyword" functionality is maintained in case authors want
%% to include these concepts in their preprints.
%%
%% You can use the \uat command to link your UAT concepts back its source.

%% From the front matter, we move on to the body of the paper.
%% Sections are demarcated by \section and \subsection, respectively.
%% Observe the use of the LaTeX \label
%% command after the \subsection to give a symbolic KEY to the
%% subsection for cross-referencing in a \ref command.
%% You can use LaTeX's \ref and \label commands to keep track of
%% cross-references to sections, equations, tables, and figures.
%% That way, if you change the order of any elements, LaTeX will
%% automatically renumber them.

\section{Introduction} 

The first discovery of gravitational waves (GWs) was a ground-breaking moment for astrophysics. The first confirmed binary black hole merger (BBH) offered new ways to understand and analyze the physics of compact objects \citep{abbott_localization_2016}. Gravitational wave multimessenger astronomy then made its own breakthrough less than a few years later thanks to GW170817 \citep{abbott_bns_2017} with its associated short gamma-ray burst \citep{Goldstein2017,Savchenko2017} and kilonova \citep[KN; e.g.,][]{Coulter2017,Troja2017,Evans2017,Arcavi2017,SoaresSantos2017,Drout2017,kasliwal_illuminating_2017,hallinan_radio_2017}. GW170817 confirmed long-standing theoretical predictions for the electromagnetic counterparts of binary neutron star mergers \citep{metzger_electromagnetic_2010,Barnes&Kasen13,metzger_kilonovae_2020} and enabled a wide range of analyses including measurements of the expansion rate of the Universe (e.g. \citealt{gw170817_nature,Hotokezaka:2018dfi,palmese_ag,2025arXiv250200239P,amsellem2025probing}) and the formation of compact object binaries (e.g. \citealt{Blanchard:2017csd,palmese_evidence_2017,Tsai:2020hpi,Kilpatrick:2021aav}). Despite the increased sensitivity of the GW detectors, no firm association of an electromagnetic transient with a gravitational counterpart has been established since 2017. Several BBH GW events have suggested counterparts \citep{Graham2020, Graham2023, Cabrera2024, 2025arXiv251020767C}, as well as BNS mergers \citep{Moroianu}, but the association with their respective GW events remains inconclusive \citep{ashton_current_2021,palmese_ligovirgo_2021,2024ApJ...977..122B,MaganaHernandez:2024yfg}.

On 2025 August 18, the LIGO-Virgo-KAGRA (LVK) Collaboration reported a low-significance gravitational-wave candidate, S250818k \citep{2025GCN.41437....1L}. Although characterized by a relatively high false alarm rate (FAR) of 2.1 per year, S250818k attracted considerable interest due to its potential association with SN 2025ulz \citep{Kasliwal2025sn, Hall2025sn, hall_at2025ulz_2025, Franz2025, OConnor2025ulz, Gillanders25ulz, Yang2025ulz, ackley_engrave_2026} a IIb supernova which may exhibit unusually red colors and a possible radio counterpart \citep{odwyer_identification_2026}. The candidate GW event was particularly notable for its sub-solar chirp mass. If astrophysical in origin, such a signal would imply the existence of a sub-solar mass neutron star, potentially requiring physics beyond standard neutron star formation channels. The “superkilonova” model has been proposed as a framework that connects rapidly rotating core-collapse scenarios with the possible formation of sub-solar mass compact objects \citep{Metzger2024, lerner_fragmentation_2025, Kasliwal2025sn}. However, given the low statistical significance of S250818k, a more robust GW detection would be required to establish a confident association.

Less than 3 months later on November 12th 2025, another sub-solar mass GW candidate, S251112cm, was reported by the LVK collaboration \citep{ligo_scientific_collaboration_ligovirgokagra_2025}. S251112cm has a FAR of 1 per 6.2 years, with a chirp mass between $0.1$ and $0.87~\text{M}_{\odot}$ and a luminosity distance of $93\pm27~\rm Mpc$. We initiated a multi-facility follow-up campaign within hours of the trigger. The Fraunhofer Telescope at Wendelstein observatory \citep[FTW; ][]{2014SPIE.9145E..2DH} was on sky within 2.4 hours, targeting in-volume galaxies identified by DESI \citep{desi_collaboration_data_2025, hall_at2025ulz_2025, gassert_ligovirgokagra_2025}. Less than 18 hours after merger, the Zwicky Transient Facility \citep[ZTF; ][]{bellm_zwicky_2018, masci_zwicky_2018, bellm_zwicky_2019, graham_zwicky_2019, masci_zwicky_2019, dekany_zwicky_2020, anand_ligovirgokagra_2025} was surveying the northern sky region of this event. The Dark Energy Camera \citep[DECam; PI: Palmese \& Andreoni; ][]{flaugher_dark_2015, hall_desi_2026} began its survey of the southern lobe of the candidate event high probability sky region 32 hours after merger. Combined, these three telescopes surveyed ${\sim}56\%$ of the probability region with at least two filters within 48 hours after merger. When combined with the Wide Field Survey Telescope \citep[WFST;][]{wang_science_2023, liu_s251112cm_2026}, the total probability covered is ${\sim}60\%$ \citep[see][]{liu_s251112cm_2026}. Another follow-up campaign for this event is presented in \citet{vieira_search_2026}, and preliminary analyses with the Vera C. Rubin LSST data are reported in \citet{macbride_ligovirgokagra_2025, anand_ligovirgokagra_2025}. 

In this work, we present the results of our search for an electromagnetic counterpart to S251112cm. In Section~\ref{sec:skn}, we overview the superkilonova scenario and estimate the maximum time delay between the onset of the supernova and the GW merger timescale, in order to motivate time windows for our counterpart searches.  In Section~\ref{sec:obs}, we layout the observing strategy for the three telescopes used to perform photometric follow up of S251112cm. In Section~\ref{sec:candidates}, we discuss the candidates identified by these programs, as well as those publicly reported, the selection methodology applied, and the spectroscopic follow-up we performed with the South African Large Telescope \citep[SALT;][]{buckley_southern_2006}, the Hobby-Eberly Telescope (HET), the Palomar 200 inch (P200), and Keck I. In particular, we report the discovery of a IIb SN, SN 2025adtq, which is within the localization volume of S251112cm. We construct a statistical model to understand when  coincident IIb's become significant associations. In Section~\ref{sec:bns}, we layout the results considering two hypothetical scenarios that could lead to an electromagnetic counterpart to the GW candidate S251112cm: the KN  and the superkilonova scenario. For the former, we use state of the art models to determine what parts of parameter space can be excluded based on the non-detection of a KN. For the superkilonova scenario, we analyze the physical parameters necessary to interpret SN 2025adtq as the counterpart to the GW candidate event. %In Section~\ref{sec:ssmm}, we discuss the possibility of two remnant sub-solar mass neutron stars merging. Then, we use SN 2025adtq to derive a possible phase space that would be required to be consistent with the current available understanding and what parameters could be derived. After that, we discuss a possible merger scenario with primordial black-holes. 
In Section~\ref{sec:ssmm}, we layout other possible formation pathways for sub-solar mass compact objects and the electromagnetic counterparts that would accompany them. Finally, in Section~\ref{sec:conclusion},  we present our conclusions and projections for future sub-solar mass events counterpart searches.

\section{Superkilonova Model and Time Delay Constraints}
\label{sec:skn}

The superkilonova (SKN) term was first coined to refer to a stellar explosion that synthesizes large quantities of $r$-process elements in the disk-ejecta following the collapse of a very massive star above the pair-instability mass gap (``collapsar''; \citealt{siegel_super-kilonovae_2022}). After S250818k and SN 2025ulz, \citet{Kasliwal2025sn} invoked the term to describe the possibility of a BNS merger that occurs in coincidence with a core-collapse supernova, either through the fission of the collapsing stellar core into two neutron stars \citep{durisen_fission_1985, imshennik_analytic_1998, Davies+02, postnov_rapidly_2016} or through the fragmentation of an accretion disk as proposed by \citet{Metzger2024,lerner_fragmentation_2025}. In part because of recent support for collapsar disk fragmentation based on first-principles hydrodynamical simulations \citep{ChenMetzger2025}, in this work we focus on the \citet{Metzger2024} scenario, as summarized below.

The disk fragmentation model requires a massive star that ends its nuclear burning evolution rotating extremely rapidly. As this likely requires either close tidal interaction or a merger with a binary companion star, the collapsing progenitor has generally had most or all of its hydrogen envelope removed, naturally predicting an association with stripped-envelope stellar explosions.  Upon collapse, the stellar core implodes to form a massive central compact object (typically a black hole--BH), surrounded by a large gaseous disk created from layers of the stellar envelope with greater angular momentum (e.g., \citealt{MacFadyen&Woosley99}).  If the gaseous disk is sufficiently massive to become gravitationally unstable and hot enough to cool efficiently through neutrinos, it can fragment into one or more bound objects \citep{PiroPfahl2007,ChenMetzger2025}, similar to models of planet formation around proto-stars. These objects undergo gravitational collapse to form NSs spanning a range of masses ${\sim} 0.01-1~M_{\odot}$ \citep{ChenMetzger2025}, including sub-solar objects otherwise forbidden to form through direct stellar core-collapse (this is due to the low electron fraction of the disk material, on which the local Chandrasekhar mass depends).

If such disk-formed NSs are born$-$or subsequently pair$-$into binaries, their subsequent mergers within the disk could generate one or more GW signals.  These are followed by at least one additional GW event as the remnant compact object(s) formed from the BNS merger(s) coalesce with the central BH.  Depending sensitively on the properties of the formed systems, such mergers are typically expected to occur with delays of hours to days after the collapse.  In addition to providing a temporal association with stripped-envelope stellar explosions, such events could in principle be accompanied by a symphony of high-energy transients, including one or more GRBs from the collapsar and BNS mergers \citep{Metzger2024}.

In assessing the likelihood of SKN candidates and motivating plausible search windows, it is critical to estimate the allowed range of delay times of the BNS merger and NS-BH mergers, relative to the stellar collapse and associated onset of the supernova explosion.  The GW inspiral time of two point masses $m_1, m_2$ on a circular orbit of initial semi-major axis $a$ can be written 
\citep{peters_gravitational_1963} 
\begin{equation}
    \label{eq:tgw}
    t_\text{GW} = \frac{5~c^5a^4}{256~G^3m_1m_2(m_1+m_2)}.
\end{equation}
Below we neglect the potential impact of gas-driven migration (e.g., \citealt{lerner_fragmentation_2025}) in either tightening or widening the initial binary system faster than GWs; however, to the extent that such a gas disk phase is typically short-lived compared to $t_{\rm GW},$ gas-migration effects can be approximately folded into uncertainties in the ``initial'' separations of the compact objects. 

Consider first a BNS merger within the disk.  For simplicity, we consider two equal mass NS (each of mass $m_\text{NS}$) in a circular orbit whose center of mass itself orbits the central black hole at a radius $r \gtrsim 100 r_{\rm g}$, where $r_{\rm g} \equiv GM_{\rm BH}/c^{2}$.  To represent a stable hierarchical triple system, the initial separation of the BNS binary, $a$, must fit tightly inside the Hill radius of the binary orbiting around the BH, 
\begin{equation}
    \label{eq:rh}
    r_{\rm H}=r\left(\frac{2m_{\rm NS}}{3M_\text{BH}}\right)^{1/3}.
\end{equation}

Taking $m_1 = m_2 = m_\text{NS}$, and scaling $a$ to half the binary Hill radius, we find from Eq.~\eqref{eq:tgw},
\begin{equation}
\begin{aligned}
\label{eq:tnsns}
t_{\rm NS-NS} ={}& 13.62\,\mathrm{hr}\,
\left(\frac{r}{300\,r_{\rm g}}\right)^4
\left(\frac{M_\text{BH}}{10\,M_\odot}\right)^{8/3} \\
&\times
\left(\frac{m_\text{NS}}{0.3\,M_\odot}\right)^{-5/3}
\left(\frac{a}{0.5\,r_{\rm H}}\right)^4 .
\end{aligned}
\end{equation}
The merger of two sub-solar NS is likely to result in the formation of a NS remnant rather than a BH \citep[e.g.,][]{Corman+26}. To estimate the time for this product NS to merge with the BH, we again use Eq.~\eqref{eq:tgw}, but this time taking $m_1 = M_\text{BH}$ and $m_2 \approx 2m_\text{NS}$, thus giving
\begin{equation}
\begin{aligned}
\label{eq:tnsbh}
t_{\rm NS-BH} ={}& 34.02\,\mathrm{hr}\,
\biggl(\frac{r}{300\,r_{\rm g}}\biggr)^4
\biggl(\frac{M_\text{BH}}{10\,M_\odot}\biggr)^3 \\
&\times
\biggl(\frac{m_\text{NS}}{0.3\,M_\odot}\biggr)^{-1}
\biggl(\frac{10.6\,M_\odot}{M_\text{BH}+2m_\text{NS}}\biggr) .
\end{aligned}
\end{equation}

Eqs.~\eqref{eq:tnsns}, \eqref{eq:tnsbh} reveal the strong sensitivity of the expected delay between the onset of the supernova and the associated GW merger times on the initial properties of the triple system, particularly the mass of the BH and its radial separation from the binary, $r$.  Nevertheless, taking fiducial estimates of $m_\text{NS} \approx 0.3M_{\odot}$ for S251112cm (Appendix~\ref{ap:compute_masses}), expected NS formation radii $r \lesssim 300 ~r_g$ \citep{Metzger2024}, and BH masses $\lesssim 20~M_\odot$ compatible with stripped-envelope SN modeling and observations \citep{schneider_pre-supernova_2021}, we find a maximum BNS delay time $t_\text{NS-NS} \sim 3.6$~days.\footnote{The allowed parameter space is constrained by the fact that the final NS-BH merger must take place significantly after the sub-solar NS mergers, i.e. $t_{\rm NS-BH} > t_{\rm NS-NS}$. If this were not the case, the NS-NS system would become tidally disrupted before merging.} We also find that increasing the NS's mass only decreases the merger time. This delay provides a maximum time window to search for supernova emission prior to sub-solar BNS mergers such as S251112cm. Motivated by this, we search for stripped envelope SNe consistent with the GW volume with estimated explosion times up to four days before the merger alert.  For these parameter values, from Eq.~\eqref{eq:tnsbh} we expect the BH-NS merger to occur within $\lesssim 11$ days of the explosion and typically a few days after the observed NS-NS merger. As we discuss in Sec.~\ref{subsubsec:parametersn2025adtq}, such a merger may well have been missed due to downtime of the LVK detectors.

%It is important to note that both timescales begin at the explosion time of the SN. This is because while the NS-NS system begins to in-spiral, the NS-NS system itself is also in-spiraling with the black hole. This is important in a situation where one detects the NS-NS merger and the final NS-BH merger along with the superkilonova. By determining the timing of all three scenarios, combined with the relative mass measurements by the LVK, one could constrain the $t_0$ of the superkilonova, allowing for the degeneracy of chance coincidence to be broken. In the case of a single detection of the NS-NS merger with a possible candidate superkilonova we can still probe the allowed parameter space to determine if a discovered SN could feasibly host a SKN.

\section{Observations}
\label{sec:obs}

\begin{figure*}[h!]
    \centering
    \includegraphics[width=1\linewidth]{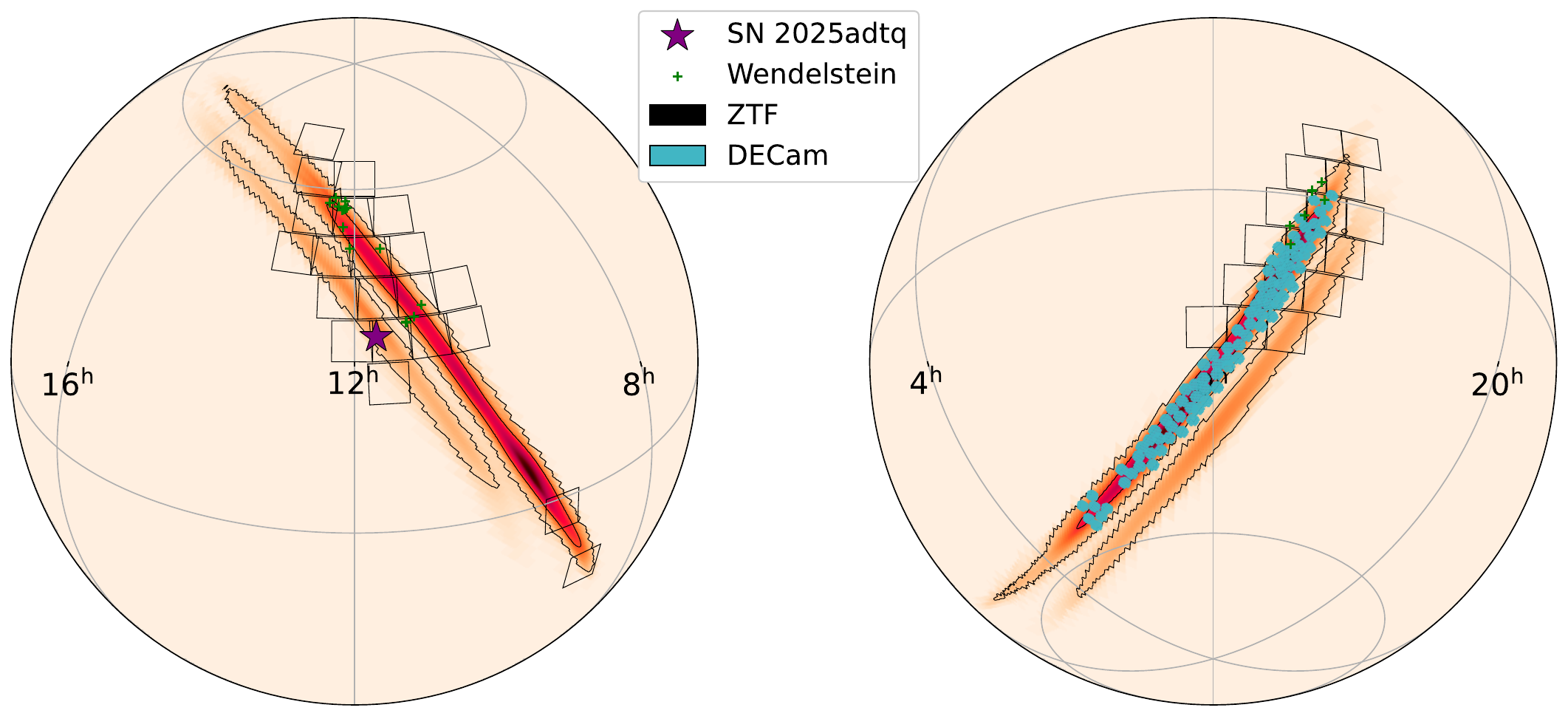}
    \caption{$50\%$ and $90\%$ credible interval sky localization of S251112cm. Overlaid are the targeted galaxy pointings by Wendelstein, the tiling performed by ZTF on the night of November 12th and the tiling performed by DECam on the night of November 13th. Some northern coverage was missed by ZTF due to poor weather. The location of the IIb SN 2025adtq is marked with a purple star.}
    \label{fig:skymap}
\end{figure*}

\begin{figure*}
    \centering
    \includegraphics[width=1\linewidth]{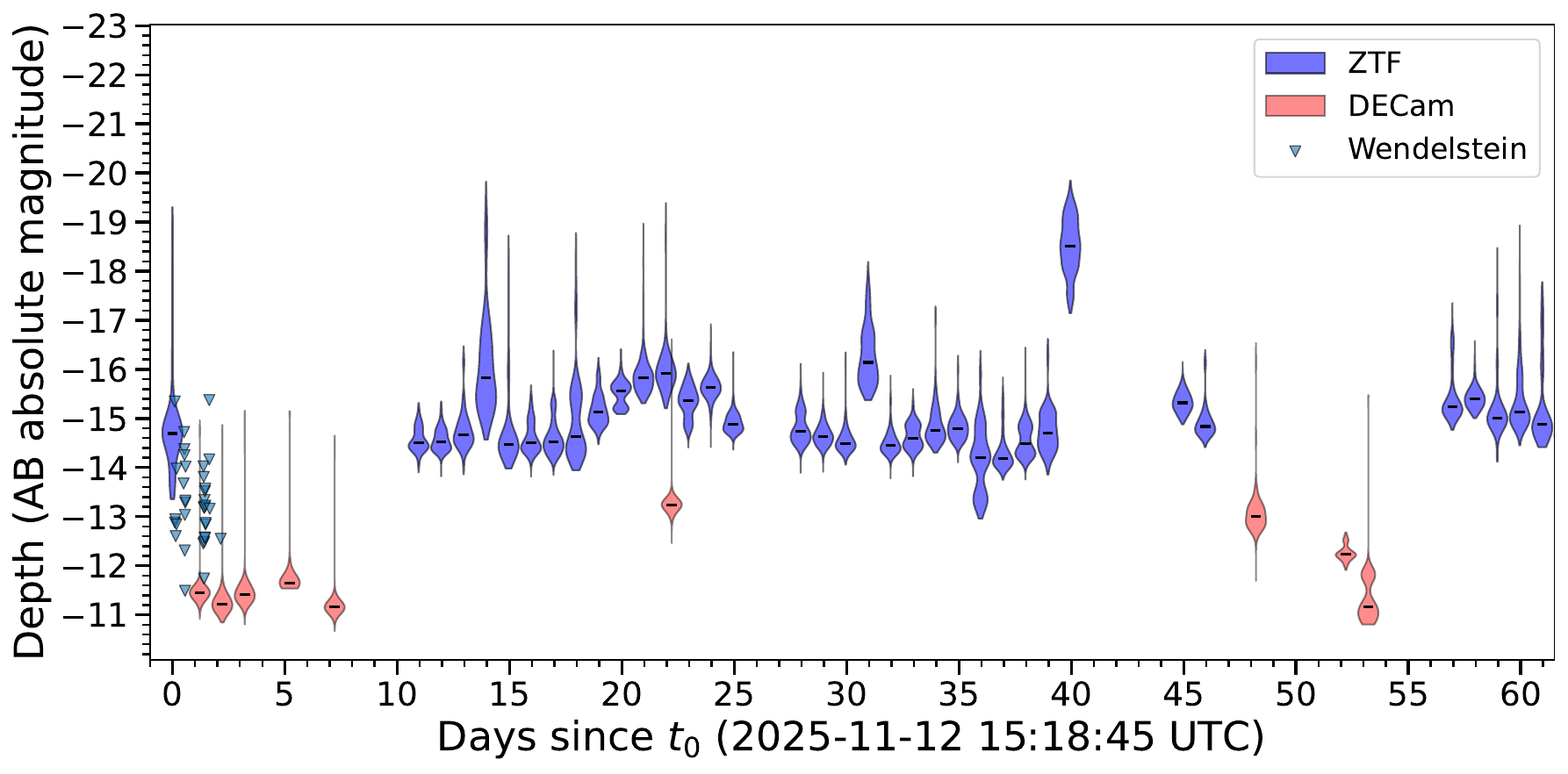}
    \caption{The ZTF and DECam g-band and Wendelstein r-band depths of the epochs over time. Absolute magnitude for ZTF and DECam is computed based on a distance of $93$~Mpc while the Wendelstein depths are computed based on the selected host's redshift.}
    \label{fig:depths}
\end{figure*}

\subsection{Galaxy-targeted Search with FTW/3KK}

Starting $2.4$ hours after the GW alert, we conducted a galaxy targeted search in the optical and near-infrared for a possible counterpart to S251112cm using the FTW's Three Channel Imager \citep[3KK; ][]{2016SPIE.9908E..44L}. We selected target galaxies from the DESI catalog \citep[see][for the publicly released data release 1, DR1]{desi_collaboration_data_2025}, following the methodology laid out in \citet{hall_at2025ulz_2025}. See Figure~\ref{fig:skymap} for the on sky position of the galaxies and Table~\ref{tab:ftw_galaxies} for the depth of each observation. The 3KK is able to image, in the optical, a $7^\prime \times 7^{\prime}$ and, in the near-infrared, an $8^\prime \times 8^{\prime}$ region in the sky. We observed in the $r$, $i$, and $J$ bands, targeting 19 different galaxies with DESI DR1 redshifts over the course of three nights, reported here (see Table~\ref{tab:ftw_galaxies}). The optical CCD and near-infrared (NIR) CMOS data were reduced using a custom pipeline \citep{2002A&A...381.1095G, 2025arXiv250314588B}. For the astrometric calibration of the images, we used the Gaia EDR3 catalog \citep{Gaia2021, 2021A&A...649A...2L, gaiaEDR3}. Tools from the AstrOmatic software suite \citep{1996A&AS..117..393B, 2006ASPC..351..112B, 2002ASPC..281..228B} were used for the coaddition of each epoch's individual exposures. We subtract the images using the Saccadic Fast Fourier Transform \citep[SFFT;][]{hu_image_2022} algorithm for image subtraction. Each co-add and subtraction were then manually inspected for transients.

\subsection{ZTF}

The Zwicky Transient Facility \citep[ZTF; ][]{bellm_zwicky_2018,graham_zwicky_2019,masci_zwicky_2019} began a search of the northern lobe and the northern part of the southern lobe $18$ hours after the alert (Figure~\ref{fig:skymap}). The initial trigger consisted of 300 s observations in the g-, and r-band, reaching median depths of $19.0$ AB mag and $20.1$ AB mag respectively. During the first night ZTF covered (i.e. observed at least twice) 38.3\% of the localization region \citep{gcn42677}. The region had been previously visited by ZTF the days prior to the GW alert, and on average, each field was observed 1.5 times in the 48 hrs prior to the alert. Due to weather conditions, ZTF could not observe the region for the next days, and only started monitoring the candidates after 11 days. Once observations commenced, and given that the region lay within the ZTF survey footprint, nightly 30 s exposures in the g and r bands were obtained over the subsequent 30 days. In addition, we triggered deeper 300 s g- and r-band exposures of the accessible region on December 18, 2025 (36 days after the GW event). By the end of the ZTF campaign, 37 days after the event, only 31\% of the region was accessible and observed. We scan transients using SkyPortal\footnote{\url{https://skyportal.io/}} \citep{walt_skyportal_2019, coughlin_data_2023}. For transients of possible interest, we used the Spectral Energy Distribution Machine (SEDM) to perform imaging follow-up \citep{blagorodnova_sed_2018, rigault_fully_2019, kim_new_2022}.
% \textcolor{red}{Tomas A. writes this.}

\subsection{DECam}

Starting $1.4$ days after the GW candidate, DECam began a wide field survey of the southern lobe of the reported Bilby skymap. Observations were delayed by a day, in order to wait for an updated offline map. With $90$, $30$s pointings in the $g$ and $i$ filters, we were able to cover $20\%$ of the total posterior probability down to median depths of $23.4$ AB mag and $22.7$ AB mag respectively. In order to scan out moving objects, we took observations for each filter at least $15$ minutes apart. In total six epochs were taken on November 13th, 14th, 15th, and 19th and December 4th and 30th with varying depths (Figure~\ref{fig:depths}). DECam observations were processed using the pipeline laid out in \citep{hu_gpu-accelerated_2026}, which uses SFFT \citep{hu_image_2022} to perform difference imaging with DECam templates. Candidates with associations to known stars in Gaia were excluded \citep{Gaia2021, 2021A&A...649A...2L, gaiaEDR3}. Furthermore, we required there to be two detections separated by ${\sim}7$ minutes to remove solar system objects. Finally, at least one of the two alerts needed to pass the real-bogus classifier cut. After this, we were left with 8092 candidates.

%\subsection{WINTER}

\section{Candidates}
\label{sec:candidates}

\subsection{Search, Selection, Vetting and Follow-up}

Due to the sub-solar nature of the merger, our scanning criteria were three-fold. First, we searched for possible transients that were discovered up to 4 days before the $t_0$ of the GW event. As discussed in Section~\ref{sec:skn}, these pre-merger transients could be SN that could host possible sub-solar mass mergers. The second criterion was some form of rapid evolution that would be consistent either with an infant SN or a KN. Finally, transients were sorted based on cross-matches to existing spectroscopic catalogs such as NED \citep{cook_ligovirgokagra_2025}, SDSS \citep{almeida_eighteenth_2023}, and DESI \citep{desi_collaboration_data_2025}.

After manual screening with these constraints, we reported 209 of the DECam transients. Each of these candidates was then vetted. We found that none of these 209 candidates appeared to have rapid evolution either consistent with that of a KN or very young SN that were also consistent with the 3D GW localization. The results of the vetting can be seen in Table~\ref{tab:decam_candidates} and Table~\ref{tab:decam_candidates_hostless}. 

%\textcolor{red}{Robert expands on ZTF scanning method.}

We search for ZTF candidates using \texttt{nuztf} \citep{bran}, the same standard framework used for past ZTF gravitational-wave counterpart searches \citep[see e.g.][]{ahumada_ligovirgokagra_2026,Kasliwal2025sn}. We perform the series of standard cuts \cite[see e.g.][]{bran}: selecting sources within the 95\% contour of the latest LVK map, requiring sources to have at least two detections separated by at least 15 minutes, rejecting sources which historical pre-detections (defined as a detection more than 4 days before the merger), and rejecting likely image artifacts or sources near bright stars. We make one modification for the `superkilonova' model: we require a time window of ZTF discovery at -4d to +3d relative to merger time, rather than the conventional 0d to +3d window. With these cuts, we find 27 ZTF candidates, including SN 2025adtq. The list is given in Table \ref{tab:ztf_candidates}.

The South African Large Telescope (SALT) was used to perform spectroscopic follow-up of nine transients that were found to be within the 2D region of S251112cm. Of these spectra, we report five Ia SNe and the redshifts of two host spectra. The results of the search can be seen in Table~\ref{tab:salt_spec}. 

Finally, we investigated any publicly classified transients on TNS that were within the $99\%$ contour. The results of this search are presented in Table~\ref{tab:tns_candidates}. Notably, SN 2025aedy is another IIb SN that is also within the volume. Using ATLAS forced photometry \citep{shingles_release_2021} and a SNCosmo model \citep{barbary_sncosmo_2025}, we estimate a $t_0$ that is $2.33\pm0.33~\text{days}$ after the GW trigger. This $t_0$ makes it incompatible with the superkilonova model as we require the SN to occur before merger time. SN 2025afhg is a Ic SN that occurred well within the $3\text{d}$ localization however using the SNCosmo model and ZTF photometry we find a $t_0$ that is $6.19\pm0.35~\text{days}$ after the GW trigger.

\subsection{SN 2025adtq}

%\begin{figure}
%    \centering
%    \includegraphics[width=0.98\linewidth]{figures/fig_ugri_baseline.png}
%    \caption{A hydrodynamical model of the lightcurve of SN 2025adtq using a combination of photometric data from WFST, ZTF, and Wendelstein. \textbf{Place holder figure}}
%    \label{fig:placeholder}
%\end{figure}

\begin{figure}
    \centering
    \includegraphics[width=0.98\linewidth]{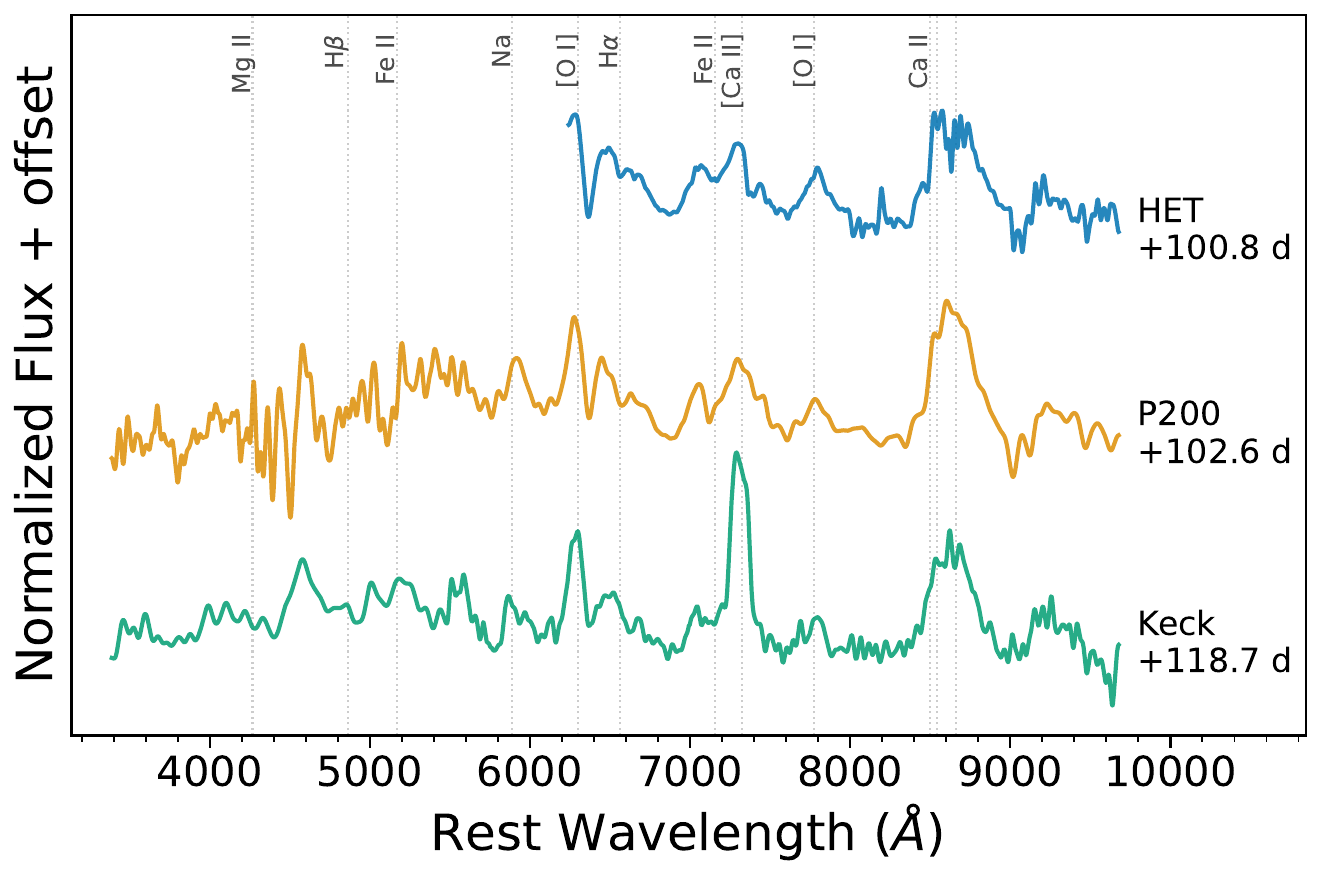}
    \caption{Late time spectral sequence of SN 2025adtq taken with HET, P200 NGPS, and Keck I LRIS. The spectra are consistent with nebular spectrum of a IIb SN.}
    \label{fig:Spec_2025adtq}
\end{figure}

One candidate of interest that survived the cuts was SN 2025adtq. The transient was offset from a nearby galaxy by $8.7^{\prime \prime}$ with a redshift from the Sloan Digital Sky Survey \citep[SDSS]{almeida_eighteenth_2023} of $z = 0.033$ which places it at $140.2$ Mpc with a S$\text{H}_0$ES cosmology \citep{riess_comprehensive_2022}. Using a combination of HET LRS, P200 NGPS and Keck LRIS spectroscopy we were able to take late time spectra ($+100 \rm~d$ post $t_0$). We use Next-Generation Superfit \citep[NGSF; ][]{goldwasser_next_2022} and find that these spectra are consistent with a nebular SN IIb spectrum (Figure~\ref{fig:Spec_2025adtq}). For the photometric lightcurve (Figure~\ref{fig:supersnec_2025adtq}), we use wide field data from ZTF and WFST paired with late time FTW.

\subsubsection{Explosion parameters from hydrodynamical light-curve modelling}
\label{subsubsec:supersnec}

\begin{figure*}
    \centering
    \includegraphics[width=1.0\linewidth]{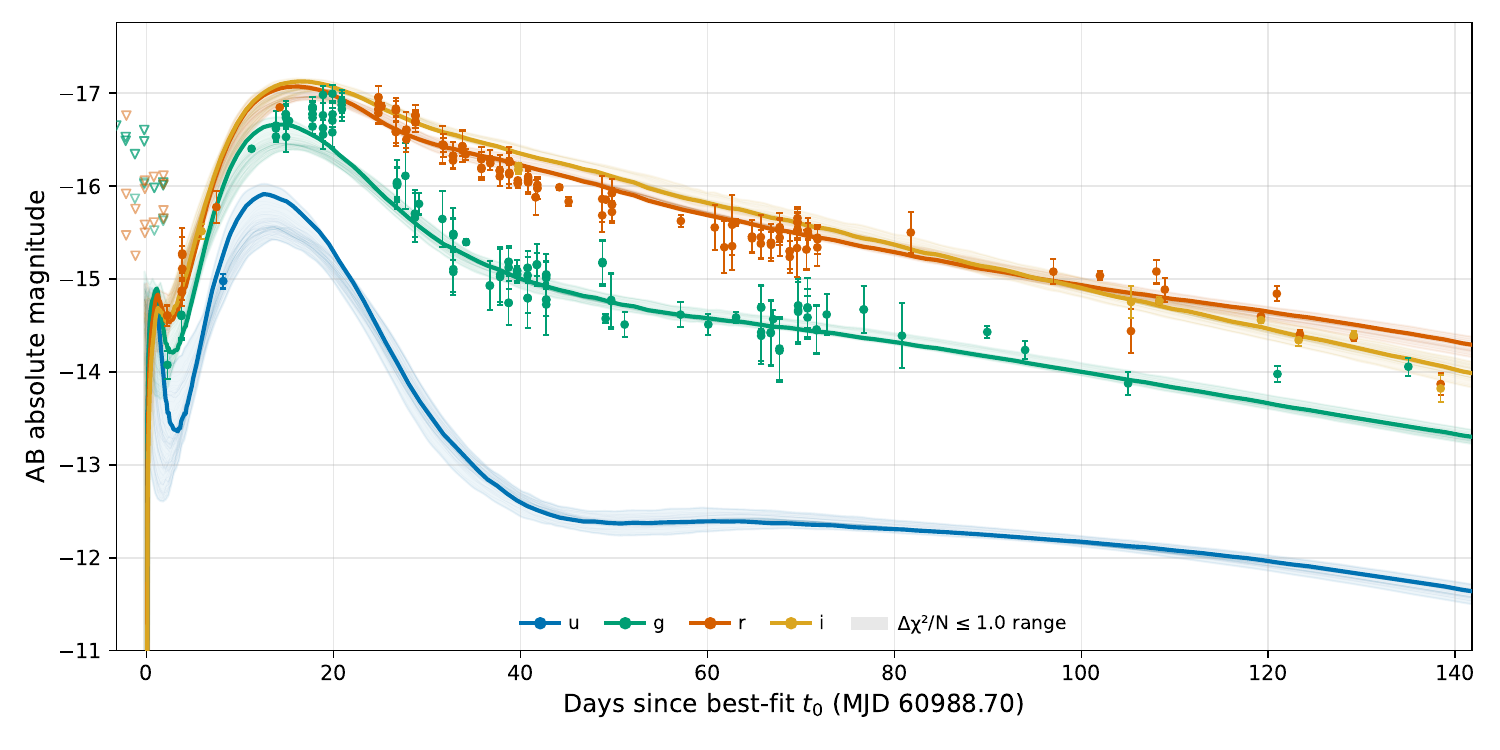}
    \caption{\texttt{SuperSNEC-lc} fit to the $u$/$g$/$r$/$i$ light curves of SN~2025adtq (Sec.~\ref{subsubsec:supersnec}). Filled circles: WFST + ZTF + ATLAS + PS1 + GOTO + Wendelstein photometry converted to absolute magnitude at $d_L = 140.2$~Mpc (WFST zero-point corrected to ZTF); inverted triangles: pre-detection $5\sigma$ upper limits. Bold coloured curve: best-fit model ($M_{\rm ej}=2.14~M_\odot$, $R_{\rm cut}=9~R_\odot$, $E_{\rm kin}=0.6\times10^{51}$~erg, $M_{\rm Ni}=0.075~M_\odot$; $t_0=\text{MJD}~60988.7$). Shaded band: envelope spanned by all models with $\Delta\chi^2/N \le 1$. Bottom panel: photospheric velocity of the best-fit model (valid only for $t\lesssim 35$~d while the 1D grey photosphere is well defined).}
    \label{fig:supersnec_2025adtq}
\end{figure*}

We model the $u$, $g$, $r$, $i$ evolution of SN~2025adtq with an in-house version of \texttt{SuperSNEC} \citep{fremling_supersnec_2026}, a derivative of the SuperNova Explosion Code \citep[\texttt{SNEC};][]{morozova_light_2015}. The modeling is based on 1D Lagrangian hydrodynamics following \texttt{SNEC}: a thermal bomb of prescribed energy is deposited at the base of a stripped pre-supernova progenitor, and the ejecta evolution is integrated through shock breakout, shock cooling, and the $^{56}$Ni-powered phase using a flux-limited diffusion radiation-transport solver on a co-moving Lagrangian grid. The input pre-supernova structure is a $13~M_\odot$ zero-age main sequence (ZAMS) helium star computed with \texttt{MESA}/\texttt{MESAstar} \citep{paxton_modules_2011,paxton_mesa_2015} from a binary system with a secondary star ZAMS mass of $11~M_\odot$ and orbital period of 50~days following \cite{Long2022}. This model has a pre-supernova radius of 60~R$_\odot$; for SN~2025adtq we further explore the effect of outer-envelope stripping by truncating the \texttt{MESA} profile at a range of trial progenitor radii $R_{\rm cut}$ before exploding it.

On top of the hydrodynamical backbone our in-house modifications of \texttt{SuperSNEC} add a layered spectral synthesis module that produces $ugriz$ synthetic magnitudes at every timestep. The synthesis combines (i) a wavelength-dependent photospheric blackbody with bandpass-specific colour-temperature dilution following \citet{dessart_quantitative_2005}, (ii) a non-thermal (NT) energy deposition treatment in which $\gamma$-rays from $^{56}$Ni/$^{56}$Co decay are channelled into ionisation, thermal heating, and forbidden/permitted line emission via a Spencer--Fano solver, and (iii) expansion-opacity line blanketing from pre-computed atomic tables for Fe, C, O, He, H and Ca. Effective opacities in Ni-rich ejecta zones are additionally modulated by a `Swiss-cheese' bubble approximation \citep{ergon_2020acat_2024}, parametrised by a maximum grey-opacity reduction $f_{\rm bub}$ that saturates above an iron-group mass fraction $x_{\rm sat}$. The spectral-synthesis layer: expansion-opacity amplitudes, the NT ionization efficiency and the wavelength-dependent dilution amplitudes were calibrated once against the well-sampled Type~IIb SN~2011dh \citep{ergon_sn2011dh_2015}, and were held fixed at their 2011dh-tuned values during the fit to 2025adtq.

To infer explosion parameters for SN~2025adtq we ran a tiered parallel grid of \texttt{SuperSNEC} models varying ejecta mass (controlled via the mass-excision coordinate, $1.5$--$2.1~M_\odot$), progenitor radius $R_{\rm cut}\in[2,60]~R_\odot$, explosion energy $E_{\rm kin}\in[0.6,1.4]\times10^{51}$~erg, $^{56}$Ni mass $0.05$--$0.11~M_\odot$, two Ni mixing parameters, two ejecta composition smoothing settings parameters, and the Ni-bubble opacity-reduction factor $f_{\rm bub}$ and its saturation $x_{\rm sat}$, leading to roughly $10^3$ independent \texttt{SuperSNEC} runs in total. Each run was scored by a $\chi^2$ metric averaged over twelve (band $\times$ epoch) cells spanning $\{g,r,i\} \times \{0$--$15, 15$--$40, 40$--$90, 90$--$140\}$~days, so that no single phase or filter dominated the residual; within each run, the explosion epoch $t_0$ was optimized in an inner scan over $[-4,+0.5]$~d relative to the first WFST detection. A small WFST to ZTF photometric zero-point offset ($+0.07$~mag in $g$, $-0.10$~mag in $r$), measured from comparing WFST data against smoothed and interpolated ZTF data, was applied to the WFST photometry before the final fit.

The best-fit model for SN~2025adtq (Figure~\ref{fig:supersnec_2025adtq}) has $M_{\rm ej}=2.14~M_\odot$, progenitor radius $R\approx 9~R_\odot$, $E_{\rm kin}=0.6\times10^{51}$~erg, synthesised $M_{\rm Ni}=0.075~M_\odot$, and explosion epoch $t_0 = \rm MJD~60988.7 \pm 1$. These values fall squarely within the range observed for Type~IIb SNe \citep{ergon_sn2011dh_2015,ergon_2020acat_2024}, and the bias-equalised $\chi^2/N = 4.75$ across 256 photometric epochs is dominated by per-epoch scatter rather than any systematic phase. The shaded band in Figure~\ref{fig:supersnec_2025adtq} shows the envelope of the 109 models with $\Delta\chi^2/N \le 1$ around the minimum, indicating that $M_{\rm ej}$, $M_{\rm Ni}$, and $t_0$ are tightly constrained by the light curve alone, while $E_{\rm kin}$ and $R_{\rm cut}$ remain modestly degenerate; the photospheric velocity of the best-fit model ($\sim 6000$~km~s$^{-1}$ near peak) is consistent with the prototypical stripped-envelope regime. Overall the photometric evolution of SN~2025adtq is fully consistent with a standard Type~IIb core-collapse supernova.

\subsubsection{Odds Ratio}

To determine if SN 2025adtq could be associated with the GW candidate S251112cm, we compute the probability that its host galaxy lies within the three-dimensional localization volume of the event. Following the formalism described in~\cite{Ashton:2017ykh}, and the same method as described in \citet{hall_at2025ulz_2025}, we use the latest and publicly available \texttt{BILBY}~\citep{Ashton:2018jfp} localization skymap~\citep{Singer_2016} encoding the posterior density over sky position and luminosity distance. Given this, we define the odds of association as given by the overlap integral defined as,

\begin{equation}
    \label{eq:odds_ratio}
    \mathcal{I} = \int 
    \frac{
        p\!\left(d_L(z_{\rm EM}) \mid \alpha_{\rm EM}, \delta_{\rm EM}\right)
        \, p(\alpha_{\rm EM}, \delta_{\rm EM})
    }{
        \pi(\alpha_{\rm EM}, \delta_{\rm EM})
    }
    \, d\alpha_{\rm EM} \, d\delta_{\rm EM} .
\end{equation}
where $\pi(\alpha_{\rm{EM}},\delta_{\rm{EM}})$ is the probability for the position of the transient to have originated from a random association and where $(\alpha_{\rm{EM}},\delta_{\rm{EM}}, z_{\rm{EM}}) $ correspond to the SN 2025adtq host galaxy position. We find $\log_{10}\mathcal{I} \approx 4.8$ with a $\text{SH}_0\text{ES}$ Cosmology \citep{riess_comprehensive_2022}. For comparison purposes, the overlap integral for GW170817 and its host galaxy NGC 4993 had $\log_{10}\mathcal{I} \approx 5.6$~\citep{Piotrzkowski:2021hhy} and the overlap integral for S250818k and SN 2025ulz are $\log_{10}\mathcal{I} \approx 4.2$ with a $\text{SH}_0\text{ES}$ Cosmology \citep{hall_at2025ulz_2025}.
As emphasized in previous work~\citep{Ashton:2020kyr,MaganaHernandez:2024yfg,2024ApJ...977..122B}, physical association cannot be determined on odds alone. A robust claim requires a physical model that explains a possible relationship between a IIb SN and sub-solar mass mergers \citep{Metzger2024, lerner_fragmentation_2025, Kasliwal2025sn}.

\subsubsection{Chance Coincidence}
\label{subsubsec:chance_coin}

Given the relative common occurrence of supernovae, we require a rigorous method to determine if such an event could have appeared by chance or if it is likely to point to a possible association. First, we compute the expected number of background SN using

\begin{equation}
    \lambda_{\rm SN} = R_{\rm SN} V_{\rm eff}\Delta t,
\end{equation}

where $R_{\rm SN}$ is the relevant volumetric SN rate, $V_{\rm eff}$ is the effective GW localization volume, and $\Delta t$ is the physically allowed coincidence window as described in Section~\ref{sec:skn}. Now, using a Poisson distribution, we define a probability of coincidence
\begin{equation}
    \label{eq:coin_one}
    P_\text{coincidence} = 1 - e^{-\lambda_{\rm SN}}% \approx \lambda_{\rm SN}.
\end{equation}

We compute the 3D volume of the minimum credible region containing SN~2025adtq's host galaxy using \texttt{ligo.skymap}'s \texttt{crossmatch} function \citep{Singer_2016}, which returns the co-moving volume searched up to and including the host's position in the GW posterior. We find $V_{\rm eff} = 4.9\times 10^{5}~\text{Mpc}^3$ with a SH0ES cosmology \citep{riess_comprehensive_2022} and $V_{\rm eff} = 6.8\times 10^{5}~\text{Mpc}^3$ with a Planck cosmology \citep{aghanim_planck_2020} cosmology. We use the IIb SN rate \citep[$0.8^{+0.6}_{-0.4}\times10^{-5}~\text{yr}^{-1}~\text{Mpc}^3~h^3_{70}$;][]{pessi_supernova_2025}. Finally, we use our defined 4 day search period in which a IIb SN could happen. With this, we find that $P_\text{coincidence} = 0.0445^{+0.0334}_{-0.0223}$ for a SH0ES cosmology and $P_\text{coincidence} = 0.0484^{+0.0363}_{-0.0242}$ for a Planck cosmology. Across both cosmological models, the $1\sigma$ rate uncertainty yields $P_{\rm coincidence} \approx 2$--$9\%$.

\subsubsection{Statistical Modeling of the Chance Coincidence}

\begin{figure*}
    \centering
    \includegraphics[width=0.49\linewidth]{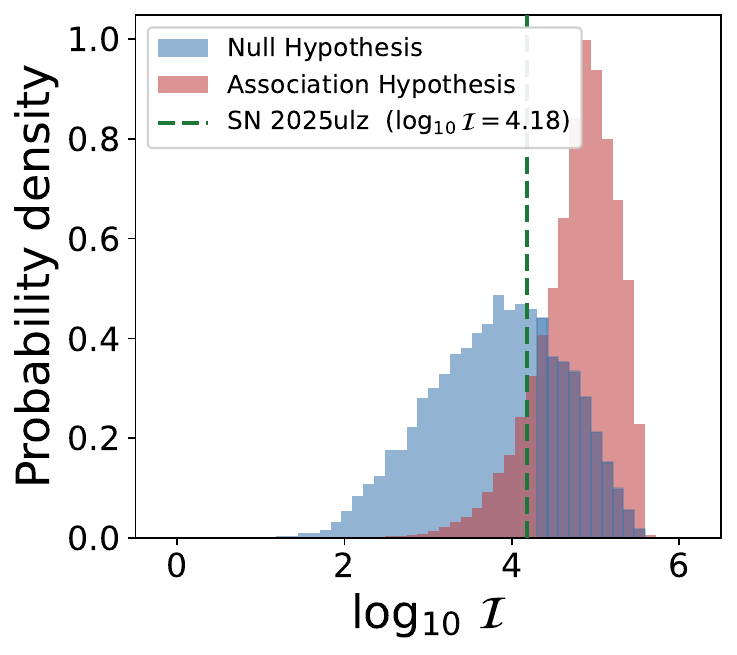}
    \includegraphics[width=0.49\linewidth]{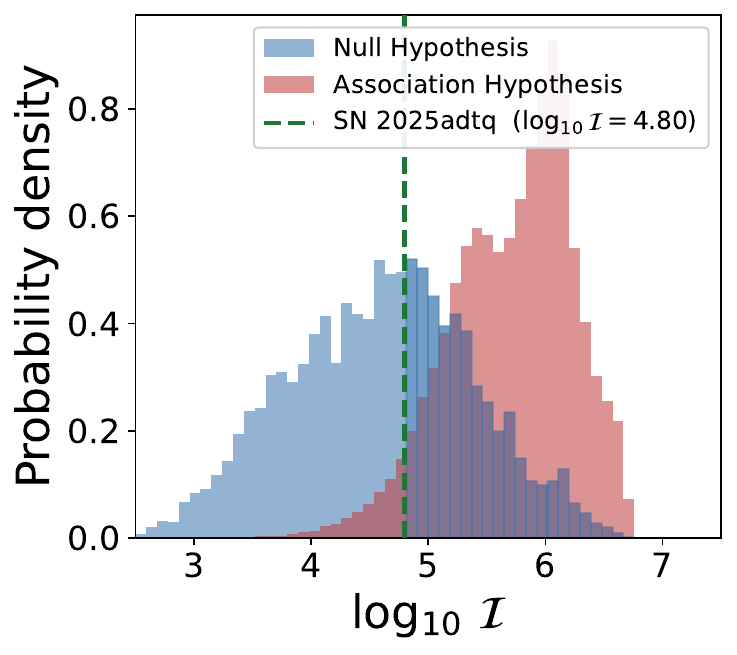}
    \caption{A comparison of the distributions of overlap integrals between two simulations, the null Hypothesis assumes that all IIb are background and records the location, while the Association Hypothesis injects a IIb SN into a random position in the 3D sky-map. The location of SN 2025ulz and 2025adtq are more consistent with expected position of a IIb from the null hypothesis when there is an association. However, this hypothesis only results in an association in $~21.3\%$ and $~4.4\%$ of scenarios respectively.}
    \label{fig:modeling_dist}
\end{figure*}

The possible in-spiral time delays of four days (computed in Section~\ref{sec:skn}), the volumetric rate of a IIb SN, and the poor localization of this event confound our ability to conclusively demonstrate an astrophysical connection between S251112cm and SN 2025adtq. If stripped envelope SNe, or a particular subtype of SN such as IIb, continues to be found within the localization volumes of  sub-solar mass merger candidates, it could possibly demonstrate the association. %We provide the following formalism to determine how many times such an association must be made in order to statistically demonstrate a relation.

In order to place into context the probability of chance coincidence, we set up a simulation that uses the \citet{pessi_supernova_2025} IIb SNe rate combined with the sky map of S251112cm. First, we consider the null hypothesis where we determine if by random chance a IIb SN would land within the 3D localization of S251112cm. To do this, we compute the number of chance coincidences expected (Sec~\ref{subsubsec:chance_coin}), from there, we randomly select a sky angle until one lands within the 2D localization. We then compute the co-moving distance bounds from $\pm3\sigma$ and use a uniform prior to place a IIb SN. Then, we consider the association-hypothesis where S251112cm was caused by a IIb. In this scenario, we inject a IIb SN randomly into the localization region of S251112cm. We run each scenario $100{,}000$ times. We then compute $\mathcal{I}$ (Equation~\ref{eq:odds_ratio}). We inspect the two distributions under the association hypothesis versus the null hypothesis and then compare them to SN 2025adtq (Figure~\ref{fig:modeling_dist}).

We define the unconditional false alarm probability (FAP) as the probability, under the null hypothesis, of obtaining a simulated IIb coincidence with an overlap statistic at least as large as that of SN~2025adtq:
\begin{equation}
\mathrm{FAP}_{\mathrm{uncond}}
=
P\!\left(
\log_{10}\mathcal{I}
\ge
\log_{10}\mathcal{I}_{\mathrm{SN\,2025adtq}}
\,\middle|\,
H_{\mathrm{null}}
\right).
\label{eq:fap_uncond}
\end{equation}

Conditioning instead on the existence of a coincident IIb SN in the localization volume, we define the conditional false alarm probability as
\begin{equation}
\begin{aligned}
\mathrm{FAP}_{\mathrm{cond}}
=
P\!\Bigl(
\log_{10}\mathcal{I}
\ge
\log_{10}\mathcal{I}_{\mathrm{SN\,2025adtq}}
\,\Big| \\
H_{\mathrm{null}},\,N_{\mathrm{coincidence}}>0
\Bigr).
\label{eq:fap_cond}
\end{aligned}
\end{equation}
where \(N_{\mathrm{coincidence}}\) is the number of simulated IIb SNe that fall within the three-dimensional localization region of S251112cm in a given trial. We find that $\mathrm{FAP}_{\mathrm{uncond}} = 1.926\%$ and $\mathrm{FAP}_{\mathrm{cond}} = 43.932\%$. Now, assuming events of comparable significance are independent,
\begin{equation} 
\label{eq:all_coin_fap}
P(C_N)=[\mathrm{FAP}]^N. 
\end{equation}

Using this, we find that, in order to reject the null hypothesis with the unconditional probabilities, that we only need 2 (5) events like S251112cm and SN 2025adtq to achieve a $3\sigma$ ($5\sigma$) level of certainty. Given the conditional probabilities, we find that we will need 12 (26) events like S251112cm and SN 2025adtq to achieve a $3\sigma$ ($5\sigma$) level of certainty. The conditional probability is significantly less dependent on the IIb SN rate and as such offers a secure upper bound of what needs to be achieved in order to qualify a statistically significant relationship between IIb SN and sub-solar mass mergers. 

We can also compute a true positive rate, which quantifies, under the association hypothesis, what percentage of associations we would expect to have a worse odds ratio than SN 2025adtq. We define this as
\begin{equation}
\mathrm{TPR}
=
P\!\left(
\log_{10}\mathcal{I}
\le
\log_{10}\mathcal{I}_{\mathrm{SN\,2025adtq}}
\,\middle|\,
H_{\mathrm{assoc}}
\right).
\label{eq:fap_uncond}
\end{equation}

For SN 2025adtq, we compute this value to be $\mathrm{TPR} = 5.14\%$. We can then take the ratio of this value against the $\mathrm{FAP}_{\mathrm{uncond}}$ to compute an odds ratio of the association hypothesis against the null hypothesis:

\begin{equation}
\Lambda
=
\frac{\mathrm{TPR}}{\mathrm{FAP}}.
\label{eq:lambda_uncond}
\end{equation}

We find that $\Lambda_{\mathrm{uncond}} = 2.669$ and $\Lambda_{\mathrm{cond}} = 0.167$. We interpret this number as determining how many more times likely $\mathrm{H}_\mathrm{assoc}$ is than $\mathrm{H}_\mathrm{null}$. Thus, the unconditional statistic provides support for an association, whereas the conditional statistic disfavors association. This disparity indicates that the apparent evidence is heavily dependent on the rarity of a coincident IIb SN occurring at all, rather than by SN~2025adtq being strongly distinguished from a chance coincidence.

\subsubsection{Statistical Significance of SN 2025ulz and SN 2025adtq}
\label{sec:ssoulzadtq}

Given that we find that we only require $N = 2$ for a $3\sigma$ relationship, it becomes clear, we should expand this methodology to S250818k and SN 2025ulz. Using the same methodology, we find that $\mathrm{FAP}_{\mathrm{uncond}} = 7.102\%$, $\mathrm{FAP}_{\mathrm{cond}} = 32.752\%$, $\mathrm{TPR} = 11.49 \%$, and $\Lambda_{\mathrm{uncond}} = 1.618$. Now we compute the joint probability using equation~\ref{eq:all_coin_fap}. We use the unconditional probabilities to find that $P(C_{2,\mathrm{uncond}}) = 7.102\% \times 1.926\% = 0.1367\%$. This means that the joint false alarm probability constitutes a $3.0\sigma$ event. We note that this calculation is sensitive to the uncertainties in the SN rate \citep{pessi_supernova_2025}, the cosmology chosen, and is premised on both gravitational wave events being real. The joint conditional probability is $P(C_{2,\,\mathrm{cond}}) = 43.932\% \times 32.752\% = 14.388\%$, indicating that, if two chance coincidences had occurred, there is a $14.388\%$ probability the data would resemble SN\,2025adtq and SN\,2025ulz. The conditional probabilities therefore do not strongly disfavor a chance coincidence interpretation on their own. To quantify the overall significance, we compute a joint $\mathrm{TPR} = 0.5906\%$, yielding a joint $\Lambda_{\mathrm{uncond}} = 4.318$ and $\Lambda_{\mathrm{cond}} = 0.041$, which reflects a preference for a true association over coincidence, but if chance coincidence has happened twice, it is consistent with chance coincidence. We caution, however, that statistical evidence for or against association does not conclusively establish the fragmentation mechanism proposed by \citet{Metzger2024}; it indicates only that IIb SNe would appear to be correlated with sub-solar mass compact binary mergers.

\section{Constraints on kilonova emission}
\label{sec:bns}

\begin{figure*}
    \centering
    \includegraphics[width=1\linewidth]{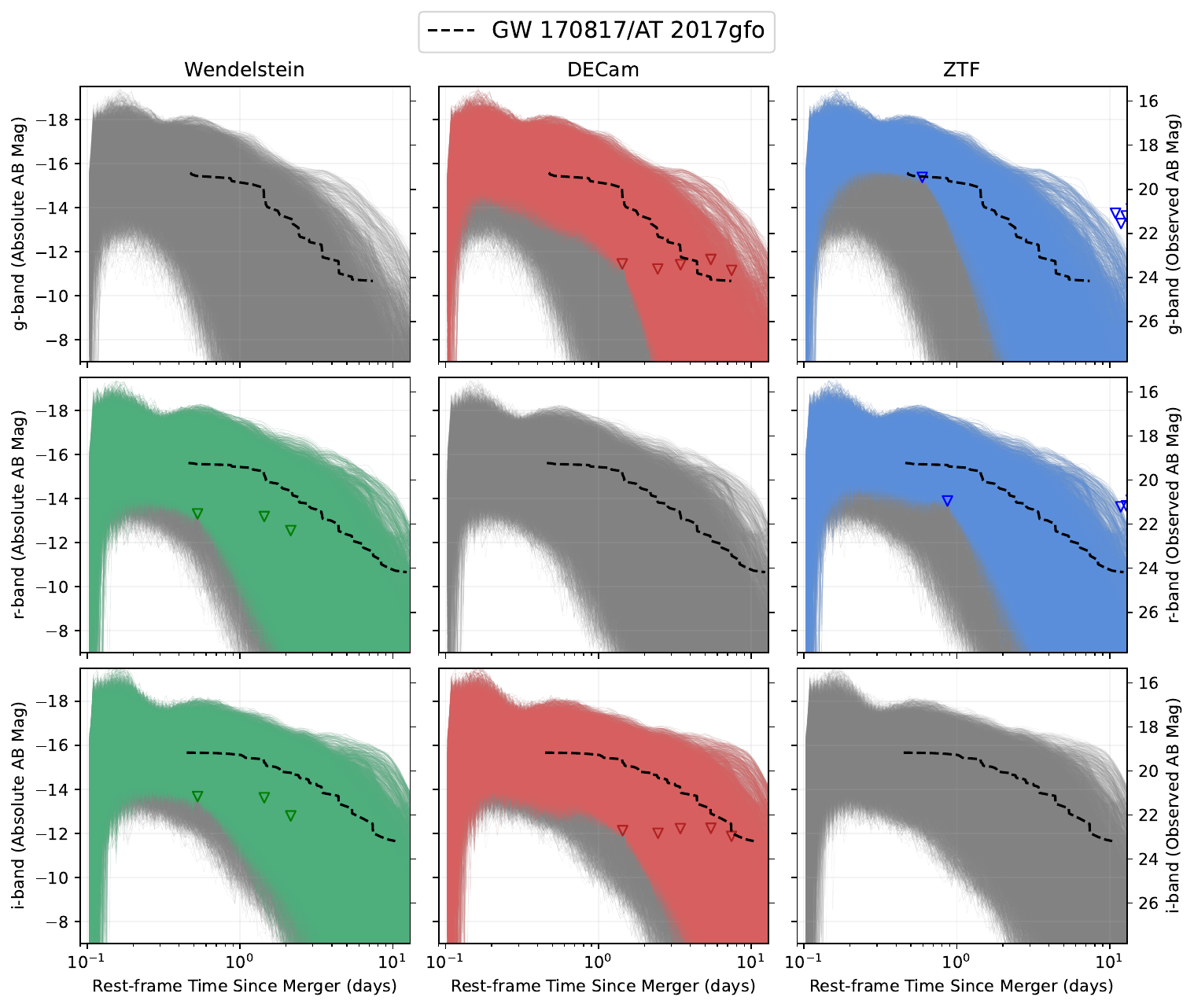}
    \caption{Kilonova light-curve simulations for BNS mergers are shown with the ZTF and DECam data scaled to a distance of 93 Mpc, while the Wendelstein data are scaled to the distance of the corresponding host galaxy. Each panel presents rest-frame synthetic light curves in a different photometric band, with observed upper limits overlaid as colored triangles. The dotted blackline represent the lightcurve of AT 2017gfo (the counterpart to GW170817) which was taken from \citet{Villar_2017}. The full model grid is shown in gray, and in each filter the models ruled out by the median observation from a given telescope are highlighted in that telescope’s corresponding color.}
    \label{fig:bns_models}
\end{figure*}

The classically considered optical counterpart of a binary neutron star merger is a KN \citep{metzger_kilonovae_2020}. Figure~\ref{fig:depths} gives the $5\sigma$ depths achieved by FTW, DECam and ZTF. We use these limits to build constraints in both coverage and depth to understand what kinds of KNe we are able to exclude as a possible optical counterpart to S251112cm, given that we did not identify any plausible KNe from our data. The lack of a KN is consistent with the findings from other surveys such as Vera C. Rubin LSST \citep{macbride_ligovirgokagra_2025, anand_ligovirgokagra_2025}, the Wide Field Survey Telescope \citep{liu_ligovirgokagra_2025, liu_s251112cm_2026}, and \citet{vieira_search_2026} which also do not yield any possible canonical KN counterparts. 

 %Furthermore, the grid is identical to that used in \citet{hu_kilonova_2025, ahumada_ligovirgokagra_2026}. Such a grid demonstrates a wide range of possible optical counterparts to a binary neutron star merger that could be related to S251112cm. These models are based on the ejecta properties from such a KN and so are still present viable model grids despite the sub-solar nature of the merger.

We use 3D radiation transfer models produced by \texttt{POSSIS} \citep{bulla_possis_2019, bulla_critical_2023}. The \texttt{POSSIS} BNS KN model grid is constructed following \citet{hu_kilonova_2025} and \citet{ahumada_ligovirgokagra_2026}, who employ the same grid in their respective follow-up campaigns for S250206dm. The grid assumes axial symmetry with two distinct ejecta components: a dynamical component ejected on dynamical timescales during the merger, and a disk wind component launched from an accretion disk formed around the merger remnant \citep{nakar_electromagnetic_2020}. Numerical relativity simulations of BNS mergers involving a sub-solar mass neutron star remain scarce; however, \citet{Corman+26} recently presented simulations of a $1.8 + 0.7~M_\odot$ system, finding dynamical ejecta masses of $M_{\rm ej,dyn} \sim 0.04~M_\odot$. These values lie above the upper bound of our dynamical ejecta grid ($M_{\rm ej,dyn} \leq 0.02~M_\odot$), suggesting that the most extreme sub-solar configurations may produce dynamical ejecta brighter than the models we constrain here. Further simulations analyzing the sub-solar mass merger scenario are expected in \citet{Dietrich_inprep}.

%As these models are based on the ejecta properties from a KN and not the intrinsic masses of the progenitors, these are still viable model grids despite the sub-solar nature of the S251112cm. 

For the dynamical ejecta, angular profiles are implemented following numerical relativity simulations, with density $\rho \propto \sin^{2}\theta$ and electron fraction $Y_{e}(\theta) \propto \cos^{2}\theta$, where $\theta$ is the polar angle with respect to the binary angular momentum axis \citep{perego_neutrino-driven_2014, radice_binary_2018, setzer_modelling_2023}. The wind ejecta are assumed to have uniform $Y_{e}$ and a spherically symmetric density distribution. Inspired by \citet{anand_chemical_2023}, the grid spans six free ejecta parameters: dynamical ejecta mass $M_{\rm dyn} = (0.001, 0.005, 0.01, 0.02)\,M_{\odot}$, mass-weighted averaged dynamical ejecta velocity $\bar{v}_{\rm dyn} = (0.12, 0.15, 0.20, 0.25)\,c$, mass-weighted averaged dynamical ejecta electron fraction $\bar{Y}_{e,\rm dyn} = (0.15, 0.20, 0.25, 0.30)$, wind ejecta mass $M_{\rm wind} = (0.01, 0.05, 0.09, 0.13)\,M_{\odot}$, mass-weighted averaged wind velocity $\bar{v}_{\rm wind} = (0.03, 0.05, 0.10, 0.15)\,c$, and wind electron fraction $Y_{e,\rm wind} = (0.20, 0.30, 0.40)$. This yields 3072 distinct model combinations, and when accounting for 11 viewing angles $\theta_{\rm obs}$ equally spaced in $\cos\theta_{\rm obs}$ from face-on ($\cos\theta_{\rm obs} = 1$) to edge-on ($\cos\theta_{\rm obs} = 0$), a total of $33{,}792$ simulated KNe are produced. %\textcolor{red}{Tomas A. can you verify this is correct? TA, yes, Mattia introduced this grid in the 0206dm paper, I cited it above}

%The upper limits for each set of observations are presented in Figure~\ref{fig:depths}. We use the median depths and compare the rest-frame times of these observations to the model lightcurves at ${\sim}93~\text{Mpc}$. Such a comparison allows us to rule out a parameter space that would be inconsistent with our observations. The results of this are shown in Figures~\ref{fig:bns_models} and \ref{fig:bns_corner}, we find that ZTF observations allow us to rule out 75.67\% of the KN models considered, Wendelstein let us rule out 96.52\%, and with DECam we rule out 92.22\% of the model grid. The probability region coverage of the first ZTF epoch ($38.3\%$) and first DECam epoch ($30\%$) means that we are able to rule out a statistically significant amount of the model grid. 

The upper limits for each set of observations are presented in Figure~\ref{fig:depths}. To derive KN model constraints, we follow the methodology of \citet{hu_kilonova_2025}. We compare the rest-frame times and median depths of our observations to the model grid lightcurves, evaluating the fraction of models brighter than our upper limits at the time of each observation (Figure~\ref{fig:bns_models}). The ZTF and DECam observations are compared to model lightcurves scaled to a fiducial distance of ${\sim}93~\text{Mpc}$ (the median of the distance posterior), while the Wendelstein observations are scaled to the distances of the individual targeted host galaxies. In the absence of constraints on the binary inclination, we account for the viewing angle dependence of the KN emission by weighting each model by the probability of its viewing angle under the prior distribution of compact binary coalescence orientations \citep{schutz_networks_2011}, which peaks at ${\sim}30^{\circ}$.

The results of this analysis are shown in Figures~\ref{fig:bns_models} and \ref{fig:bns_corner}. We find that ZTF observations allow us to rule out $42\%$ of the KN models considered, Wendelstein rules out $92\%$, and DECam rules out $68\%$ of the model grid. The dominant dependence is on the $M_{\rm wind}$. Models with larger wind ejecta masses are most strongly excluded across all three data sets, while models with $M_{\rm wind}=0.01\,M_{\odot}$ make up the bulk of the unconstrained models. It is important to note that these rule-out parameters are based on being above the limits from both filters from a single epoch. The probability region covered by the first ZTF epoch is $43\%$ and first DECam epoch is $20\%$. Combining the total region observed by ZTF $(43\%)$ and DECam $(20\%$) with an overlap of $5.6\%$. Doing a weighted average of the percentage each telescope ruled out and its coverage percentage allows us to conclude a ${\sim}29\%$ probability that the combined observations rule out a canonical KN scenario. % and combined ($??\%$) means that we are able to rule out a statistically significant amount of the model grid.

\section{Applying the fragmentation superkilonova model}
\label{sec:applyskn}

\begin{figure*}
    \centering
    \includegraphics[width=1\linewidth]{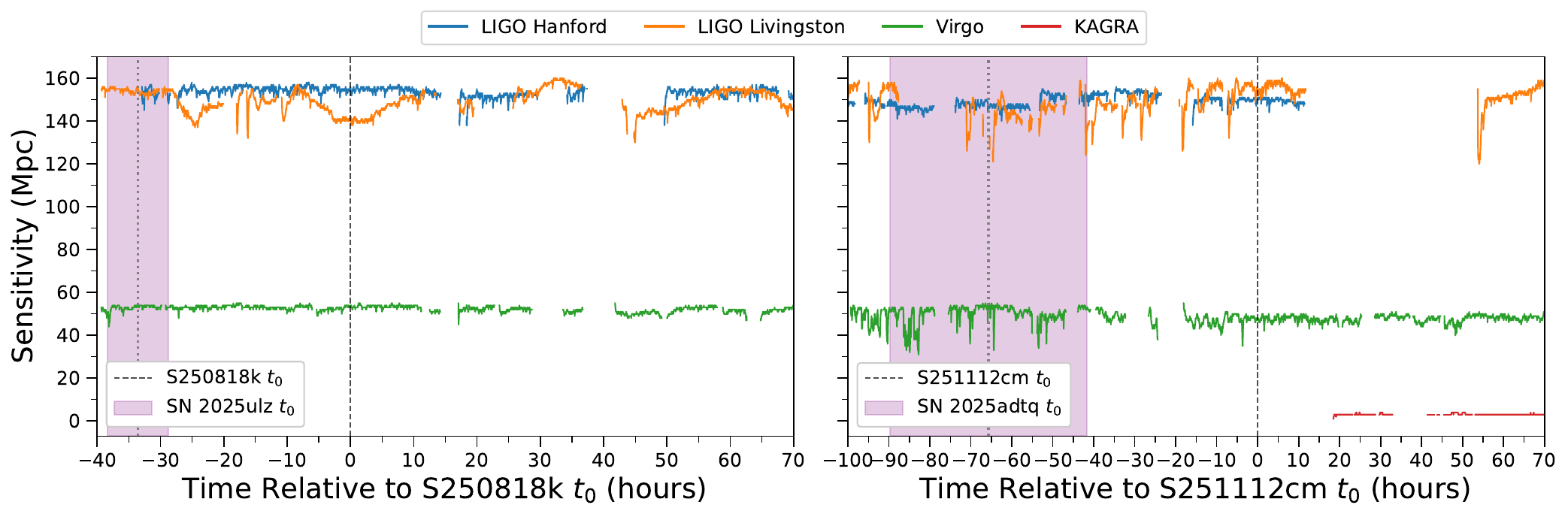}
    \caption{Sensitivity of the GW detectors at a given time. \textbf{Left:} It can be seen that less than two days after S250818k, the Lingston and Hanford observatories were down for around 6 hours. \textbf{Right:} It can be seen that less than one day after S251112cm, the Livingston and Hanford observatories were not on for about 42 hours. In both cases, when these detectors were off, they would not detect the possible NS-BH merger that would come after the sub-solar mass merger.}
    \label{fig:ligo_downtime}
\end{figure*}

The fragmentation superkilonova model (\citealt{Metzger2024}; Sec.~\ref{sec:skn}) offers an possible formation channel for sub-solar mass mergers. It also offers a possible electromagnetic counterpart in the form of a stripped envelope SN. Unfortunately, the latter are relatively common and as such it will require substantive evidence to prove such association (Sec.~\ref{sec:ssoulzadtq}). The possibility that SN 2025adtq could be such a superkilonova is worth exploring. SN 2025adtq is now the second such example of a IIb SN that is consistent with the localization of a sub-solar mass merger candidate \citep{Kasliwal2025sn, hall_at2025ulz_2025, Hall2025sn, Franz2025, Gillanders25ulz, odwyer_identification_2026}.

\subsection{The Parameter Space of SN 2025adtq}
\label{subsubsec:parametersn2025adtq}

Conclusive evidence of the fragmented superkilonova scenario could come from a detection of $r$-process signatures or an associated long- or short-GRB (potentially detectable in X-ray observations or late-time radio as found by \citealt{odwyer_identification_2026}). We note that the r-process emission may also be buried under the supernova emission. However, the most conclusive evidence of such a scenario (regardless of an optical counterpart) would be a spatially and temporally coincident sub-solar mass BNS merger and NS-BH merger. Unfortunately, less than ${\sim}1$ day after S251112cm the Hanford and Livingston Observatories turned off for ${\sim}2$ days which is around the time when according to the model (Equation~\ref{eq:tnsbh}) a NS-BH merger could be expected (Figure~\ref{fig:ligo_downtime}).

For SN 2025adtq, the $\sim$2 day window during which the Hanford and Livingston observatories were offline, at the expected time for the NS-BH merger, provides a non-detection constraint that we can use as a consistency check on the model parameters. We derive an explosion time estimate of SN 2025adtq ($t_{0,\rm SN} = 60988.7 \pm 1~\text{d}$) from our fit in section~\ref{subsubsec:parametersn2025adtq}. We then find a feasible parameter space, derived using the equations in Section~\ref{sec:skn}, with a $6$--$20~\text{M}_\odot$ BH and a formation radius of $250$--$560~r_\text{g}$ (Figure~\ref{fig:inspiral}). These values fall within typical limits on a IIb BH remnant being less than $20~\text{M}_\odot$ and around our fiducial estimate of $300~r_\text{g}$ \citep{schneider_pre-supernova_2021}. While this internal consistency is encouraging for the superkilonova scenario, we emphasize that observing a NS-BH merger would provide the most confident evidence for the superkilonova scenario. %is neither confirmed nor excluded by these constraints; the parameter space derived here is consistent with the model, but consistency alone does not constitute evidence for it

\subsection{The Parameter Space of SN 2025ulz}

\begin{figure*}
    \centering
    \includegraphics[width=0.98\linewidth]{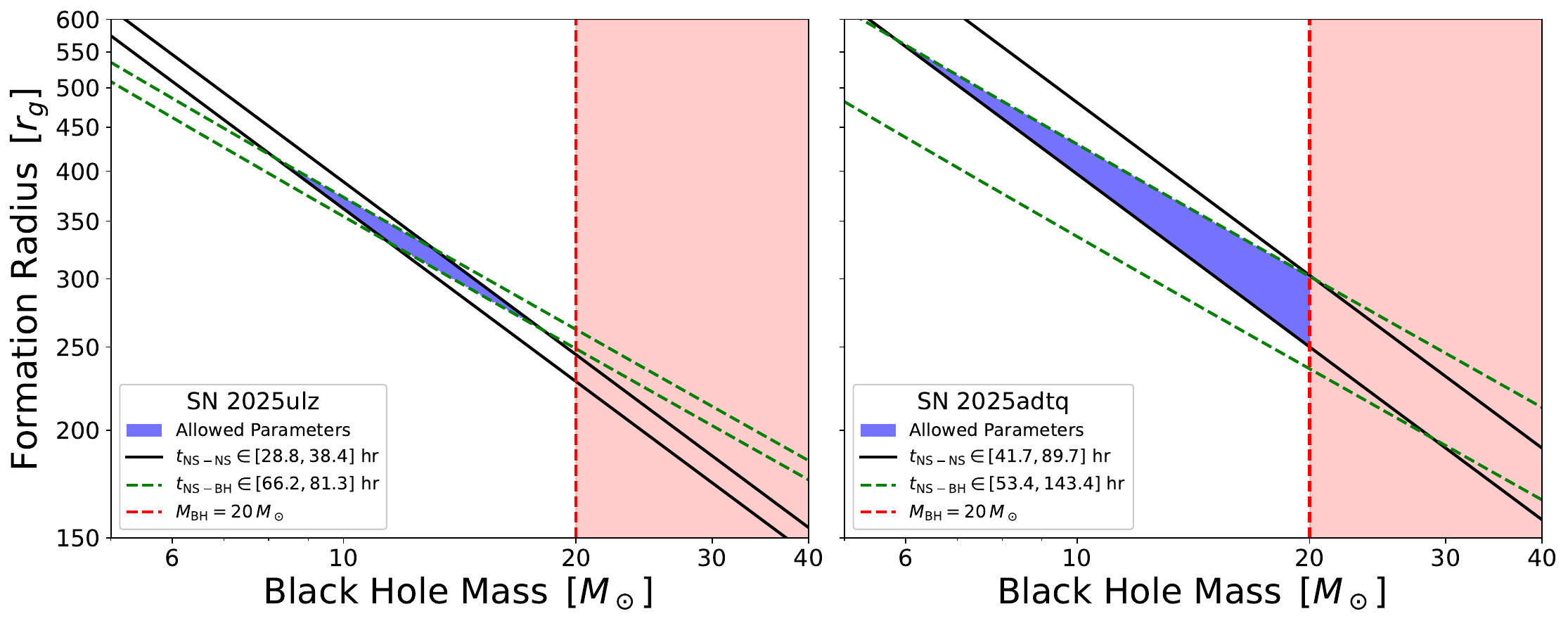}
    \caption{Allowed parameter space (purple region) of the mass of the BH remnant $(M_\text{BH})$ and the formation radius of a putative sub-solar mass NS binary in the BH accretion disk, in units of gravitational radii, $r_{\rm g} \equiv GM_{\rm BH}/c^{2}$. The region between the solid lines show corresponds to setting the inspiral time of the NS-NS binary (Eq.~\eqref{eq:tnsns}) equal to the estimated explosion time $t_0$ of SN 2025adtq with respect to the merger time of S251112cm.  We also demand that the subsequent BH-NS merger would have been missed, which requires that the inspiral time (Eq.~\eqref{eq:tnsbh}) fall within the downtime window of the LIGO detectors, as denoted by the region between the green dashed lines.  We also impose an upper limit on $M_{\rm BH}$ of $20~M_\odot$ (vertical dashed line) based on the expected compact objects formed from IIb SNe.  These three constraints allow for a significant parameter space in BH mass ($6$--$40~\text{M}_\odot$) and formation radius ($250$--$450~r_\text{g}$) consistent with the superkilonova scenario for SN 2025adtq/S251112cm. We also perform the same analysis on SN 2025ulz and S250818k and find an equally permissible parameter space.}
    \label{fig:inspiral}
\end{figure*}

We can repeat the same exercise for SN 2025ulz as there was a smaller $6$ hour window during which both Hanford and Livingston observatories were offline also around the expected time for a NS-BH merger. We take the explosion time estimate of SN 2025ulz from \citet{Kasliwal2025sn} which is $t_{0,\rm SN} - t_{0,\rm GW}  = 1.4\pm 0.2~\text{d}$. From this, we again find a parameter space with a $8$--$18~\text{M}_\odot$ BH and a formation radius of $250$--$450~r_\text{g}$ (Figure~\ref{fig:inspiral}). This is once again encouraging for the possibility of the superkilonova model, however a detection of an actual NS-BH merger will provide the most confident evidence for a superkilonova scenario.

\subsection{Reconciling LIGO non-detections of other IIb SNe}

If a statistically significant relationship between IIb SNe and sub-solar mergers can be demonstrated, this raises the issue of how to distinguish IIb SNe that appear related to sub-solar merger GW events versus those that have no temporally coincident GW event, and how often the fragmentation superkilonovae scenario occurs. Not including SN 2025ulz and SN 2025adtq, during the general on-period of the LIGO detectors during O4, there have been 73 IIb SNe classified on TNS within $400~\rm Mpc$ and 58 IIb SNe within $150~\rm Mpc$. However, LIGO's sensitivity to sub-solar mass mergers is substantially reduced relative to standard BNS, BHNS, or BBH events, meaning a significant fraction of these events may have produced GW signals below detection threshold even during nominal observing periods. A detailed offline search of O4 for sub-solar events \citep[such as][]{kacanja_search_2026, ligossmassmerger} or even a directed search towards publicly classified IIb SNe \citep[similar to what was done with the galactic center; ][]{abbott_search_2022} could therefore prove essential in determining whether other sub-solar mass mergers may have been hosted by IIb SNe within LIGO's effective sensitivity range.

The fragmentation superkilonova scenario itself requires very specific conditions such as high angular momentum, significant in-fall onto the BH, and finally fragmentation and formation of the sub-solar mass neutron stars (Sec.~\ref{sec:skn}). Such necessary conditions explain why the scenario may be rare in general, but offer little guidance in understanding why it might be rare specifically among IIb progenitors. IIb SNe likely have multiple progenitor channels, not all of which involve binary interaction that substantially spins up the stellar core just prior to core collapse. The fact that SN~2025ulz appeared redder than the standard IIb population \citep{Kasliwal2025sn} while SN~2025adtq appears photometrically standard (Appendix~\ref{ap:iib_pop}) suggests that color alone is not a reliable discriminant between the two classes. To fully address this, some other observable property distinguishing the standard IIb from the superkilonova IIb must be identified to break the current degeneracy. A detailed population study of IIb SNe, examining properties such as hydrogen envelope mass, explosion energy, ejecta velocity, remnant properties, and progenitor system geometry, will be necessary to determine whether the hypothesized superkilonova IIb progenitors occupy a distinct region of this parameter space.

The photometric normality

\section{Other Electromagnetic Counterparts to Sub-solar Mass Mergers}
\label{sec:ssmm}

Well-measured neutron star masses all exceed a solar mass (e.g., \citealt{Lattimer21}), broadly consistent with the predictions of modern core-collapse supernova simulations (e.g., \citealt{Burrows&Vartanyan21,Janka25}).  It is non-trivial to form sub-solar compact objects in nature, much less for them to find another compact object with which to merge. Here, we discuss a few other possibilities and their associated electromagnetic counterparts.

\subsection{Sub-solar remnants of phase transitions }

\citet{essick_exotic_2024} highlights that equations of state (EoS) with large, high-density phase transitions can produce exotic stable branches at sub-solar masses, so-called High-pressure Objects (HiPOs). In this framework, when a normal NS accretes sufficient mass to reach its TOV maximum, rather than collapsing to a black hole it instead transitions to the denser, stable HiPO branch. This is analogous to how a white dwarf exceeding the Chandrasekhar limit collapses onto the NS branch. Here the transition drives the ejection of as much as $\mathcal{O}(1)~M_\odot$ of material. Because the ejected material would be highly neutron-rich, the resulting electromagnetic transient is expected to more closely resemble a KN due to significant expected $r$-processing. As a result of the large ejecta mass and high energy, such an event could be expected to be as bright as a superluminous SN. No such event has been confidently observed to date, and the rate is expected to be low as not all NSs are likely to undergo this process. Such a formation process is relevant here as at least one of the objects in S251112cm is of a sub-solar mass. As such one could imagine a formation channel of a binary system that results in a more standard neutron star that eventually merges with the HiPO. However, \citet{essick_exotic_2024} notes that sub-solar mass HiPO objects possess tidal deformabilities of $\Lambda \sim \mathcal{O}(10)$, which are small enough that they could be confused with sub-solar mass black holes ($\Lambda_\mathrm{BH} = 0$) in GW observations, complicating the astrophysical interpretation of any such detection.

\subsection{Sub-solar mass black holes}

Given the fact that the equations of state show that a sub-solar mass neutron star would be indistinguishable from a black hole to LIGO \citep{essick_exotic_2024}, we must also consider the possibility that either one or both of the objects observed in this merger candidate are primordial black holes (PBHs). Sub-solar mass black holes cannot be produced through standard stellar evolution, and a well-motivated formation channel for compact objects in this mass range is through primordial density fluctuations in the early universe \citep{prunier_analysis_2024, kacanja_search_2026}. 

In the case of a PBH-NS merger, \citet{markin_general-relativistic_2023} present detailed relativistic hydrodynamical simulations of a merger between a $0.5~M_\odot$ BH with a $1.4~M_\odot$, which has a chirp mass (Equation~\ref{eq:chirp}) of $0.71 ~M_\odot$ and would be feasibly consistent with the binned chirp mass available for S251112cm. Using the \texttt{POSSIS} \citep{bulla_possis_2019, bulla_critical_2023} code, they predict that an EM counterpart would peak at $M_g \sim -14$ ($M_i \sim -15$) and then would decay rapidly, falling below $M_g \sim-8$ ($M_i \sim -12$) within two days. Such a candidate would be consistent with a KN and as such should be picked up by existing searches. %Comparing such a model with the depths achieved in our search campaign (Figure~\ref{fig:depths}), such a model could have had single detections on the first epoch observations of each ZTF, DECam, and Wendelstein, but would almost certainly be gone by any second epoch. No transient observed has a rapid decent akin to this, and, if one was observed, it would certainly have been considered a feasible KN candidate.

Some scenarios attempt to connect PBH mergers to either fast radio bursts (FRBs) or gamma ray bursts (GRBs) \citep{deng_fast_2018, liu_merger_2020}. Since there has been no reported overlap of FRBs or GRBs with S251112cm, these scenarios remain unconfirmed. If there is a lack of EM counterparts, the other possibility is that the merger is effectively a BBH coalescence and that there may not be an EM counterpart. \citet{riajul_haque_primordial_2026} specifically analyze S251112cm under a PBH binary interpretation and find that it is viable within current observational bounds.%, with merger probabilities reaching unity in the $0.5$--$1~M_\odot$ mass range.% If the fragmentation superkilonova scenario is rejected and we see continued non-detections of KNe or PBH-NS mergers, this may give evidence that these sub-solar mass events are binary PBH mergers.

\section{Conclusions}
\label{sec:discussion}
\label{sec:conclusion}
We present the analysis of our combined DECam, ZTF, Wendelstein, SALT, HET, P200, and Keck follow-up for an electromagnetic counterpart to S251112cm, a candidate sub-solar mass compact object merger. Considering a KN counterpart to S251112cm, our dataset rules out $42$--$92\%$ of canonical KN emission models over $56.4\%$ of the total sky localization. As such, we rule out a combined $29\%$ of the sky and model grid. Without full coverage of the entire localization and without all KN models ruled out, a canonical KN counterpart to S251112cm may have been missed.

Complementary wide-field coverage of S251112cm obtained with the 2.5-m Wide Field Survey Telescope \citep[WFST;][]{wang_science_2023}, a facility designed in part for rapid target-of-opportunity follow-up of GW triggers \citep{liu_target--opportunity_2023} and which has already delivered competitive KN constraints on earlier candidates such as S250206dm \citep{liu_illuminating_2026}, will be presented here in a forthcoming joint analysis \citep{liu_s251112cm_2026}. Only through a coordinated combination of wide-field surveys with complementary depth, cadence, and sky coverage can place competitive constraints on canonical KN emission across the full localization volume of an event like S251112cm.

To understand how we should pursue follow-up of sub-solar merger events we discuss various possible interpretations of the merger alongside their theorized electromagnetic counterparts. We first discuss the feasibility of a canonical KN scenario and use HiPOs to offer a feasible formation channel for a more classical BNS scenario \citep{essick_exotic_2024}. From there, we investigate the PBH-NS merger scenario and determine that the theorized counterparts would either require a canonical KN-like, fast radio burst, or gamma ray burst counterpart. Finally, we flesh out the formal theory of the in-spiral times of the fragmentation superkilonova scenario determining that we should search for SN counterparts with explosion times up to $4$ days before a sub-solar merger.

We report the discovery of a IIb SN (SN 2025adtq) that is consistent with the localization and timing of S251112cm. Next, we determine that in order to demonstrate a statistically significant association between a IIb SN and sub-solar mass mergers (assuming a S251112cm and SN 2025adtq-like scenario), using the unconditional false alarm probability, we require at least 2 events to establish a $3\sigma$ significance and $5$ events to achieve a $5\sigma$ result. We perform statistical modeling and find that SN 2025ulz and SN 2025adtq together give ${\sim}3\sigma$ evidence against the null hypothesis. The unconditional odds ratio of the association hypothesis ($\mathrm{H}_\mathrm{assoc}$) and the null hypothesis ($\mathrm{H}_\mathrm{null}$) is $4.318$. This points to the probabilties favoring $\mathrm{H}_\mathrm{assoc}$ over $\mathrm{H}_\mathrm{null}$. However, we find that the joint conditional odds ratio is $0.041$. This discrepancy arises because, once we condition the probabilities on there being a chance coincidence, the 3d localizations become more consistent with $\mathrm{H}_\mathrm{null}$ than with $\mathrm{H}_\mathrm{assoc}$, even though the unconditional odds favor an association. Under the condition that there is always a chance coincidence, SN 2025adtq and SN 2025ulz are more consistent with the chance coincidence hypothesis than the association hypothesis. Eventually, if the association is real, it will be possible to establish it using the same formalism as in \citet{palmese_ligovirgo_2021}, and even constrain cosmological parameters accounting for the uncertainty of association \citep{2024PhRvD.110h3005B}.

A notable difference between the superkilonova candidates is their photometric contrast: SN~2025ulz is reported to exhibit anomalously red colors interpreted as a possible early-time $r$-process contribution to the optical SED \citep{Kasliwal2025sn}, while SN~2025adtq is photometrically standard (Appendix~\ref{ap:iib_pop}). If both associations are determined to be real, color is not a reliable discriminant, as the $r$-process ejecta in SN~2025adtq-like events may simply be buried beneath the SN photosphere at early epochs. As such, late-time nebular and infrared spectroscopy with \textit{JWST} could probe the innermost ejecta for $r$-process enrichment signatures that could show evidence for disk fragmentation and sub-solar mass NS formation \citep{Metzger2024}.

If astrophysical in origin, S251112cm is a first of a kind GW candidate that provides the strongest evidence to date for sub-solar compact objects that would require the invocation of new physics. Offline searches have also begun looking for possibly missed sub-solar mass mergers in past LVK data \citep{kacanja_search_2026}. Given the FAR of S251112cm, the lack of a discovered KN counterparts throughout most of the GW sky localization, our competitive constraints on possible KN models, and the existence of a IIb SN within the localization volume, the fragmentation superkilonova model offers a tantalizing explanation for this GW event candidate. SN 2025adtq combined with SN 2025ulz begins to portray an intriguing possibility that IIb SN could be associated to sub-solar mass mergers; however, this joint statistical evidence is sensitive to the IIb SN volumetric rate uncertainty \citep{pessi_supernova_2025} and the choice of cosmology. The superkilonova interpretation rests on several underlying assumptions, first this entire analysis is predicated on an astrophysical origin of S251112cm, secondly our maximum time delay of $4$ days is premised on fiducial model parameters such as the NSs masses, the BH mass, and the formation radius. Finally, the chance coincidence computation relies heavily on the IIb SN rate from \citet{pessi_supernova_2025} and the chosen cosmological parameters \citep{riess_comprehensive_2022}.

In the case of S251112cm, the ability to confirm or reject a superkilonova scenario was significantly hampered by detector downtime after the event (Figure~\ref{fig:ligo_downtime}) which constitutes a gap in coverage rather than a true observational non-detection. Despite non-detection of a NS-BH counterpart, application of the fragmentation superkilonova model to SN~2025adtq yields a consistent parameter space of BH remnant mass $6$--$40\,M_\odot$ and formation radii of $250$--$550\,r_{\rm g}$, broadly within the limits expected from IIb SN progenitors \citep{schneider_pre-supernova_2021}. We note that this parameter space is derived from fiducial estimates of the BH mass, NS formation radius, and sub-solar NS mass, and neglects gas-driven migration \citep{lerner_fragmentation_2025} which could modify the inferred inspiral timescales. However, confirming this scenario ultimately requires a detection of the subsequent NS-BH merger. In the future, continued data collection for several days after a sub-solar mass merger should be considered critical to determine the viability of the fragmentation superkilonova model. Likewise, we strongly suggest an offline directed GW search towards other classified IIb SNe for both sub-solar mass mergers and NS-BH mergers.

As we move into the era of wide-field survey follow-up programs such as the Vera C. Rubin's ToO program \citep{andreoni_rubin_2024} and the Argus Array \citep{law_low-cost_2022}, we will be better able to probe the sky before more of these sub-solar mass mergers are discovered. This will be critical to the superkilonova model allowing for better constraints on infant stripped envelope SN that could be the possible hosts of these GW events. Critically, comprehensive multi-wavelength follow-up of candidate superkilonova stripped envelope SNe will be essential to distinguish them from the standard IIb population. As we move into the LVK intermediate runs (IR) and eventually onto the fifth observing run (O5), the observational aspects of the merger delay time of the superkilonova model (Equations~\ref{eq:tnsns} and \ref{eq:tnsbh}), the determination of the statistical significance (Equation~\ref{eq:all_coin_fap} and \ref{eq:lambda_uncond}) of these associations, and the reconciliation of the superkilonova with standard IIb SNe, will prove essential in determining the long-term feasibility of the model.

%Equation of state modeling supports the idea that accreting neutron stars could enter a regime of immense mass loss resulting in a sub-solar mass neutron star being left behind \citep{essick_exotic_2024}.

%% Please use the acknowledgment and contribution environments. This will 
%% be anonomyized when the "anonymous" style option is used. 
\begin{acknowledgments}

X.J.H. thanks Armin Rest for his useful insight in considering gravitational wave candidates from other IIb SN. X.J.H thanks Emily McPike for useful comments on the manuscript.

A. P. is supported by NSF Grant No. 2308193. B. O. is supported by the McWilliams Postdoctoral Fellowship in the McWilliams Center for Cosmology and Astrophysics at Carnegie Mellon University. M. B. is supported by a Student Grant from the Wübben Stiftung Wissenschaft. B.l M is supported by NASA (grant 80NSSC26K0299) and the National Science Foundation (grant AST-2406637). The Flatiron Institute is supported by the Simons Foundation.

This paper contains data obtained at the Wendelstein Observatory of the Ludwig-Maximilians University Munich. We thank Christoph Ries, Michael Schmidt and Silona Wilke for performing the observations. Funded by the Deutsche Forschungsgemeinschaft (DFG, German Research Foundation) under Germany's Excellence Strategy – EXC-2094/2 – 390783311.

This work used resources on the Vera Cluster at the Pittsburgh Supercomputing Center (PSC). Vera is a dedicated cluster for the McWilliams Center for Cosmology and Astrophysics at Carnegie Mellon University. We thank the PSC staff for their support of the Vera Cluster.

Based on observations at Cerro Tololo Inter-American Observatory, NSF’s NOIRLab (NOIRLab Prop. ID 2023B-851374, PI: Andreoni \& Palmese), which is managed by the Association of Universities for Research in Astronomy (AURA) under a cooperative agreement with the National Science Foundation. We thank Kathy Vivas, Alfredo Zenteno, and CTIO staff for their support with DECam observations.

This project used data obtained with the Dark Energy Camera (DECam), which was constructed by the Dark Energy Survey (DES) collaboration.
Funding for the DES Projects has been provided by the US Department of Energy, the US National Science Foundation, the Ministry of Science and Education of Spain, the Science and Technology Facilities Council of the United Kingdom, the Higher Education Funding Council for England, the National Center for Supercomputing Applications at the University of Illinois at Urbana-Champaign, the Kavli Institute for Cosmological Physics at the University of Chicago, Center for Cosmology and Astro-Particle Physics at the Ohio State University, the Mitchell Institute for Fundamental Physics and Astronomy at Texas A\&M University, Financiadora de Estudos e Projetos, Fundação Carlos Chagas Filho de Amparo à Pesquisa do Estado do Rio de Janeiro, Conselho Nacional de Desenvolvimento Científico e Tecnológico and the Ministério da Ciência, Tecnologia e Inovação, the Deutsche Forschungsgemeinschaft and the Collaborating Institutions in the Dark Energy Survey.

The Collaborating Institutions are Argonne National Laboratory, the University of California at Santa Cruz, the University of Cambridge, Centro de Investigaciones En\`ergeticas, Medioambientales y Tecnol\`ogicas–Madrid, the University of Chicago, University College London, the DES-Brazil Consortium, the University of Edinburgh, the Eidgenössische Technische Hochschule (ETH) Zürich, Fermi National Accelerator Laboratory, the University of Illinois at Urbana-Champaign, the Institut de Ci\'encies de l’Espai (IEEC/CSIC), the Institut de F\'isica d’Altes Energies, Lawrence Berkeley National Laboratory, the Ludwig-Maximilians Universit\:at M\:unchen and the associated Excellence Cluster Universe, the University of Michigan, NSF’s NOIRLab, the University of Nottingham, the Ohio State University, the OzDES Membership Consortium, the University of Pennsylvania, the University of Portsmouth, SLAC National Accelerator Laboratory, Stanford University, the University of Sussex, and Texas A\&M University.

Some of the observations reported in this paper were obtained with the Southern African Large Telescope (SALT).

This research used data obtained with the Dark Energy Spectroscopic Instrument (DESI). DESI construction and operations is managed by the Lawrence Berkeley National Laboratory. This material is based upon work supported by the U.S. Department of Energy, Office of Science, Office of High-Energy Physics, under Contract No. DE–AC02–05CH11231, and by the National Energy Research Scientific Computing Center, a DOE Office of Science User Facility under the same contract. Additional support for DESI was provided by the U.S. National Science Foundation (NSF), Division of Astronomical Sciences under Contract No. AST-0950945 to the NSF’s National Optical-Infrared Astronomy Research Laboratory; the Science and Technology Facilities Council of the United Kingdom; the Gordon and Betty Moore Foundation; the Heising-Simons Foundation; the French Alternative Energies and Atomic Energy Commission (CEA); the National Council of Humanities, Science and Technology of Mexico (CONAHCYT); the Ministry of Science and Innovation of Spain (MICINN), and by the DESI Member Institutions: \url{www.desi.lbl.gov/collaborating-institutions}. The DESI collaboration is honored to be permitted to conduct scientific research on I’oligam Du’ag (Kitt Peak), a mountain with particular significance to the Tohono O’odham Nation. Any opinions, findings, and conclusions or recommendations expressed in this material are those of the author(s) and do not necessarily reflect the views of the U.S. National Science Foundation, the U.S. Department of Energy, or any of the listed funding agencies.

Based in part on observations obtained with the Hobby-Eberly Telescope (HET), which is a joint project of the University of Texas at Austin, the Pennsylvania State University, Ludwig-Maximillians-Universitaet Muenchen, and Georg-August Universitaet Goettingen. The HET is named in honor of its principal benefactors, William P. Hobby and Robert E. Eberly. We acknowledge the Texas Advanced Computing Center (TACC) at The University of Texas at Austin for providing high performance computing, visualization, and storage resources that have contributed to the results reported within this paper. The Low Resolution Spectrograph 2 (LRS2) was developed and funded by the University of Texas at Austin McDonald Observatory and Department of Astronomy, and by Pennsylvania State University. We thank Sergey Rostopchin, Amy Westfall, Cassie Crowe, and Justen Pautzke from the HET staff for obtaining these observations. We thank the Leibniz-Institut fur Astrophysik Potsdam (AIP) and the Institut fur Astrophysik Goettingen (IAG) for their contributions to the construction of the integral field units. We would like to acknowledge that the HET is built on Indigenous land. Moreover, we would like to acknowledge and pay our respects to the Carrizo \& Comecrudo, Coahuiltecan, Caddo, Tonkawa, Comanche, Lipan Apache, Alabama-Coushatta, Kickapoo, Tigua Pueblo, and all the American Indian and Indigenous Peoples and communities who have been or have become a part of these lands and territories in Texas, here on Turtle Island.

Based on observations obtained with the Samuel Oschin Telescope 48-inch and the 60-inch Telescope at the Palomar Observatory as part of the Zwicky Transient Facility project. ZTF is supported by the National Science Foundation under Award \#2407588 and a partnership including Caltech, USA; Caltech/IPAC, USA; University of Maryland, USA; University of California, Berkeley, USA; Cornell University, USA; Drexel University, USA; University of North Carolina at Chapel Hill, USA; Institute of Science and Technology, Austria; National Central University, Taiwan, and the German Center for Astrophysics (DZA), Germany. Operations are conducted by Caltech's Optical Observatory (COO), Caltech/IPAC, and the University of Washington at Seattle, USA. 

The ZTF forced-photometry service was funded under the Heising-Simons Foundation grant \#12540303 (PI: Graham). 

SED Machine is based upon work supported by the National Science Foundation under Grant No. 1106171 

The Gordon and Betty Moore Foundation, through both the Data-Driven Investigator Program and a dedicated grant, provided critical funding for SkyPortal. 

Some of the data presented herein were obtained at Keck Observatory, which is a private 501(c)3 non-profit organization operated as a scientific partnership among the California Institute of Technology, the University of California, and the National Aeronautics and Space Administration. The Observatory was made possible by the generous financial support of the W. M. Keck Foundation. Some of the data presented herein were obtained at Keck Observatory, which is a private 501(c)3 non-profit organization operated as a scientific partnership among the California Institute of Technology, the University of California, and the National Aeronautics and Space Administration. The Observatory was made possible by the generous financial support of the W. M. Keck Foundation. 

The Liverpool Telescope is operated on the island of La Palma by Liverpool John Moores University in the Spanish Observatorio del Roque de los Muchachos of the Instituto de Astrofisica de Canarias with financial support from the UK Science and Technology Facilities Council.

B.D.M. acknowledges support from NASA (80NSSC26K0299), the National Science Foundation (AST-2406637), and the Simons Foundation (727700).  The Flatiron Institute is supported by the Simons Foundation.

M.B. acknowledges the Department of Physics and Earth Science of the University of Ferrara for the financial support through the FIRD 2025 grant

M.W.C. acknowledges support from the National Science Foundation with grant numbers PHY-2117997, PHY-2308862 and PHY-2409481.

\end{acknowledgments}

%% To help institutions obtain information on the effectiveness of their 
%% telescopes the AAS Journals has created a group of keywords for telescope 
%% facilities.
%
%% Following the acknowledgments section, use the following syntax and the
%% \facility{} or \facilities{} macros to list the keywords of facilities used 
%% in the research for the paper.  Each keyword is check against the master 
%% list during copy editing.  Individual instruments can be provided in 
%% parentheses, after the keyword, but they are not verified.
\facilities{WO:2m (3kk), PO:1.2m (ZTF), PO:1.5m (SEDM), Hale (NGPS), Keck:I (LRIS), HET, SALT (RSS), WFST:2.5m }

%% Similar to \facility{}, there is the optional \software command to allow 
%% authors a place to specify which programs were used during the creation of 
%% the manuscript. Authors should list each code and include either a
%% citation or url to the code inside ()s when available.
\software{Astropy \citep{2013A&A...558A..33A,2018AJ....156..123A,2022ApJ...935..167A},  
          Cloudy \citep{2013RMxAA..49..137F}, 
          Source Extractor \citep{1996A&AS..117..393B}
          }

%% Appendix material should be preceded with a single \appendix command.
%% There should be a \section command for each appendix. Mark appendix
%% subsections with the same markup you use in the main body of the paper.
%%
%% Each Appendix (indicated with \section) will be lettered A, B, C, etc.
%% The equation counter will reset when it encounters the \appendix
%% command and will number appendix equations (A1), (A2), etc. The
%% Figure and Table counter will not reset.

\appendix

\section{Estimating Individual Masses}
\label{ap:compute_masses}

While LIGO does not publicly release individual masses, it does release a chirp mass binning which is defined as

\begin{equation}
    \label{eq:chirp}
    \mathcal{M} = \frac{(m_1 m_2)^{3/5}}{(m_1 + m_2)^{1/5}}.
\end{equation}

\begin{figure*}
    \centering
    \includegraphics[width=\linewidth]{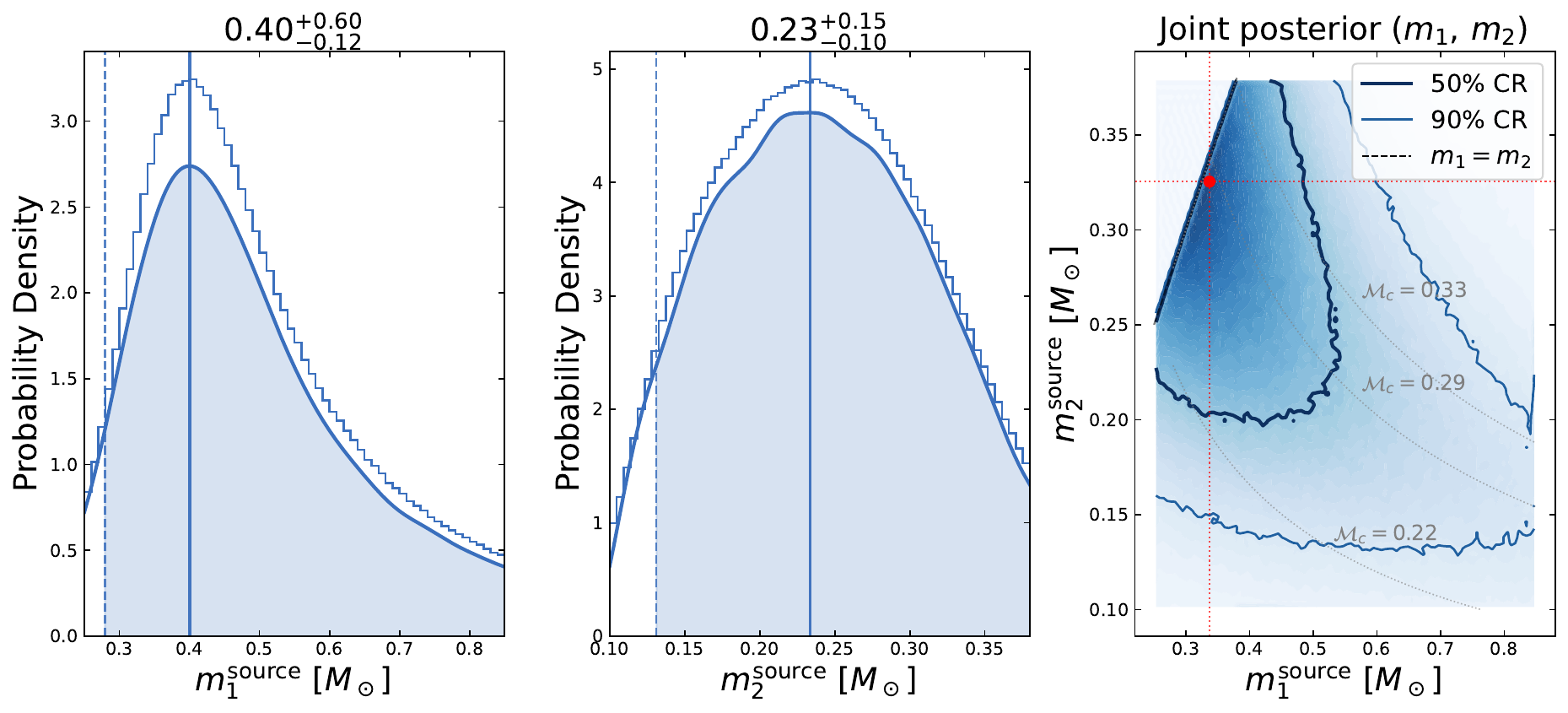}
    \caption{A posterior plot of feasible masses for $m_1$ (left)  and $m_2$ (middle) given a low chirp mass of $0.295\pm0.06$ and sampled over a uniform prior of $q \in [0.1, 1]$. Vertical lines indicate the mode values of the posteriors for $m_1$ and $m_2$. The rightmost plot is a combined posterior curve with the red value being the mode probability.}
    \label{fig:bns_masses}
\end{figure*}

Given that we know $m_1$ and $m_2$ have some posterior distribution and thus their errors would be folded into any computation of the chirp mass, the fact that the chirp mass bin is $100\%$ between $0.1$ and $0.87~M_\odot$ allows us to determine some information regarding the binary merger. The absence of any probability in the $0.87$ to $1~M_\odot$ bin is itself informative, given a high-significance event with chirp mass near $0.87~M_\odot$ would be expected to have some probability above this boundary. As such, we adopt a Gaussian prior on $\mathcal{M}_c$ centered on the geometric mean of the bin edges, $\mathcal{M}_c^* = \sqrt{0.10 \times 0.87} \approx 0.295~M_\odot$, as a best guess on the mean for a multiplicative mass parameter. We choose a $1\sigma$ width of $0.06~M_\odot$ and a uniform prior on the mass ratio $q \in [0.1,1.0]$. This is consistent with the probability bins reported by LIGO. The results of the Markov chain Monte Carlo (MCMC) of this result is shown in Figure~\ref{fig:bns_masses}. This gives $m_1 = 0.40^{+0.60}_{-0.12}~M_\odot$ and $m_2 = 0.23^{+0.15}_{-0.1}~M_\odot$ which provides us with our fiducial estimate of $m_1=m_2\simeq0.3~\text{M}_\odot$. We note that these values are sensitive to the priors chosen; however, both posteriors are consistent with ${\sim}0.3~M_\odot$.

\section{Kilonova Emission constraints}

Here we demonstrate the model parameter space that was excluded by the observations of ZTF, DECam, and Wendelstein (Figure~\ref{fig:bns_corner}).

% Page 1
\begin{figure*}
    \centering
    \includegraphics[width=\linewidth]{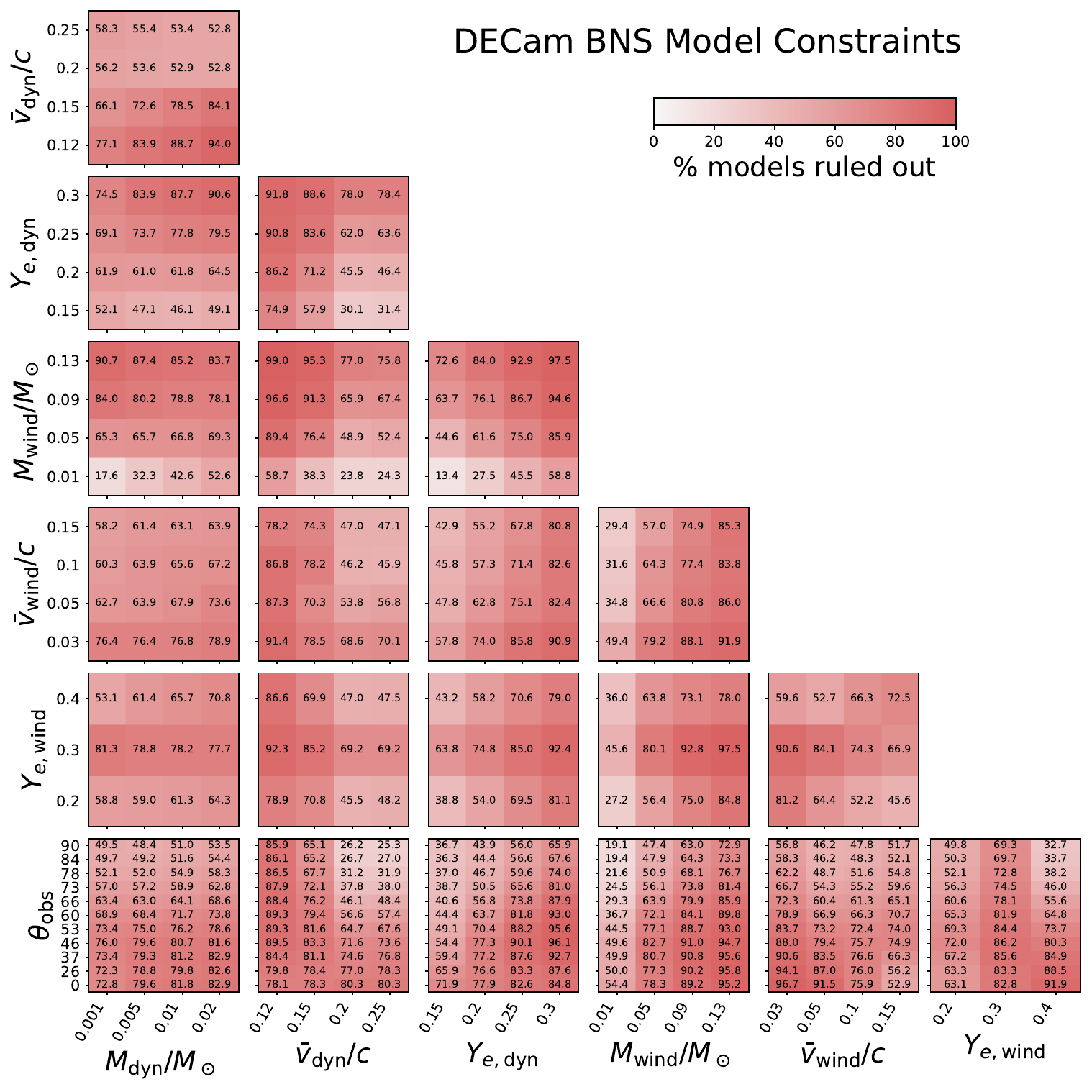}
    \caption{Corner plot showing the fraction of BNS models ruled out by our observations. Each cell is labeled by the percentage of models that were able to be ruled out with those parameters.}
    \label{fig:bns_corner}
\end{figure*}

% Page 2
\begin{figure*}
    \ContinuedFloat
    \centering
    \includegraphics[width=\linewidth]{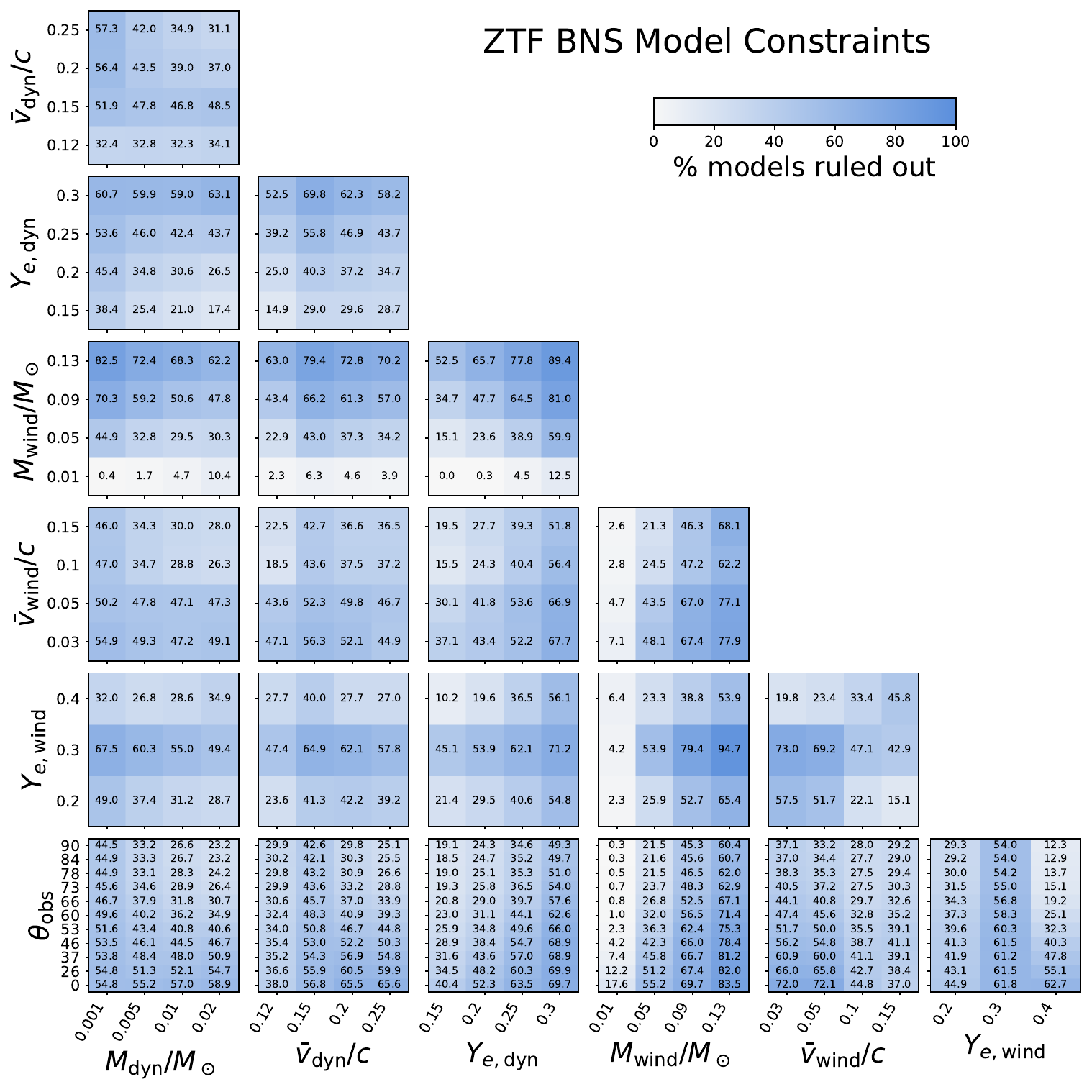}
    \caption[]{Corner plot showing the fraction of BNS models ruled out by our observations (continued).}
\end{figure*}

% Page 3
\begin{figure*}
    \ContinuedFloat
    \centering
    \includegraphics[width=\linewidth]{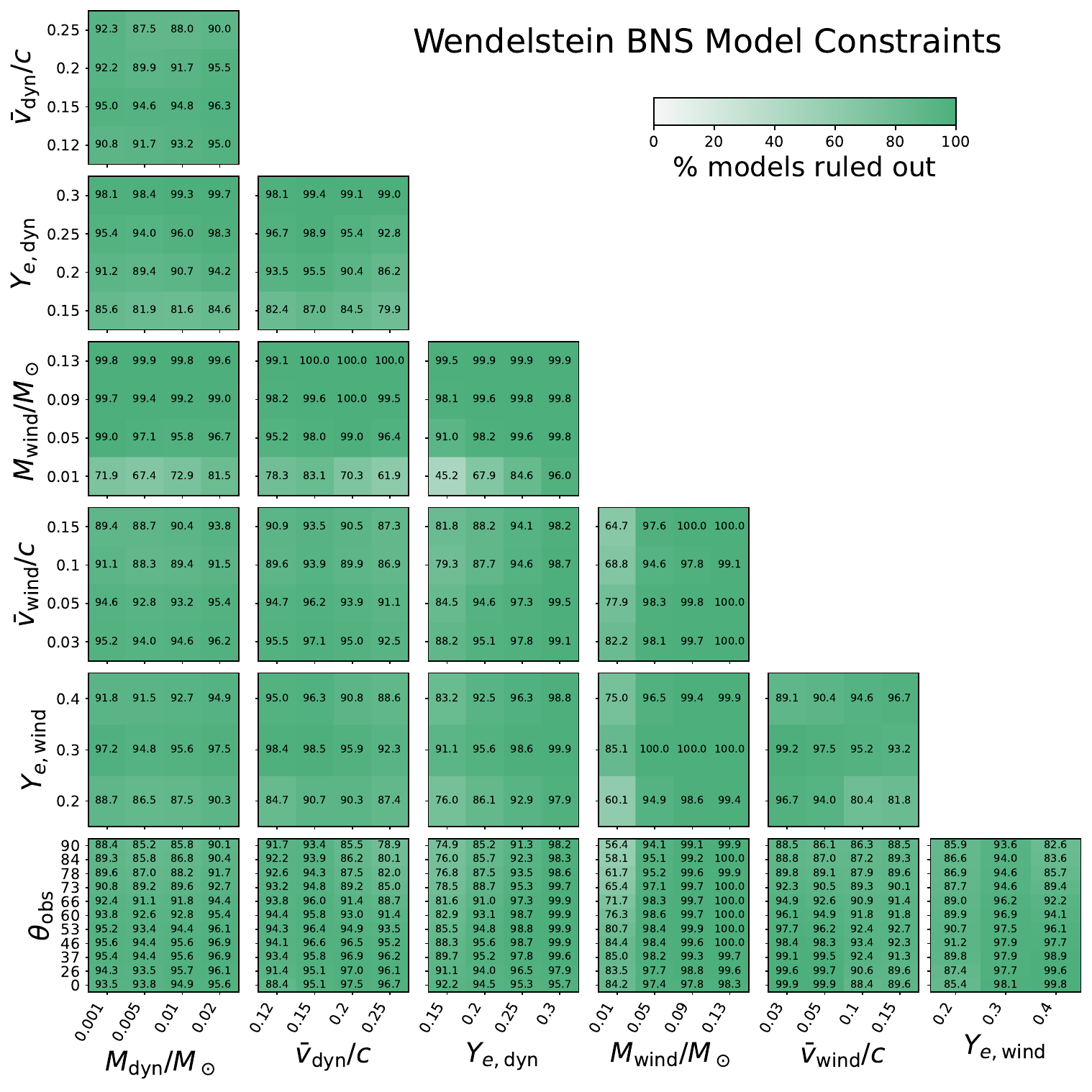}
    \caption[]{Corner plot showing the fraction of BNS models ruled out by our observations (continued).}
\end{figure*}

\section{Targeted Follow-up of In-Volume Host Galaxies}

Here, in Table~\ref{tab:ftw_galaxies}, we present the galaxies selected from the DESI catalog \citep{desi_collaboration_data_2025} that were followed up by the 2.1m Fraunhofer Telescope at Wendelstein Observatory.

\begin{table}
\caption{A table of all the in-volume galaxies observed by Wendelstein. \label{tab:ftw_galaxies}}
\begin{tabular}{c|c|c|c|c|c|c}
Name & RA [deg] & Dec [deg] & $M_{\mathrm{lim},r}$ & $M_{\mathrm{lim},i}$ & $M_{\mathrm{lim},J}$ & $t-t_0 \mathrm[d]$ \\
\hline
%S251112cm\_231853\_m101533 & 349.719 & -10.259 & -15.344 & -15.349 & -17.749 & 0.1023 \\
S251112cm\_224424\_m000943 & 341.102 & -0.162 & -12.879 & -12.985 & -15.473 & 0.1131 \\
S251112cm\_224619\_p031236 & 341.581 & 3.210 & -12.947 & -13.169 & -15.399 & 0.1255 \\
%S251112cm\_230009\_m124828 & 345.037 & -12.808 & -12.603 & -12.751 & -15.294 & 0.1481 \\
S251112cm\_225802\_m034611 & 344.509 & -3.770 & -12.846 & -12.999 & -15.147 & 0.1782 \\
S251112cm\_225249\_p011204 & 343.205 & 1.201 & -13.978 & -14.055 & -16.215 & 0.1857 \\
%S251112cm\_164320\_p703757 & 250.834 & 70.633 & -13.677 & -13.866 & -15.990 & 0.4971 \\
%S251112cm\_105742\_p373916 & 164.427 & 37.655 & -14.724 & -15.082 & -16.980 & 0.5229 \\
S251112cm\_110955\_p365648 & 167.481 & 36.947 & -14.374 & -14.714 & -16.534 & 0.5323 \\
S251112cm\_110206\_p384714 & 165.524 & 38.787 & -14.253 & -14.485 & -16.394 & 0.5398 \\
S251112cm\_121133\_p574415 & 182.889 & 57.737 & -12.310 & -12.486 & -14.772 & 0.5475 \\
S251112cm\_120945\_p563125 & 182.439 & 56.524 & -13.036 & -13.225 & -15.453 & 0.5550 \\
S251112cm\_122402\_p582307 & 186.009 & 58.385 & -11.491 & -11.667 & -13.879 & 0.5625 \\
S251112cm\_113415\_p490235 & 173.561 & 49.043 & -13.288 & -13.553 & -15.651 & 0.5702 \\
%S251112cm\_130847\_p621618 & 197.194 & 62.272 & -13.325 & -13.806 & -15.449 & 0.5778 \\
%S251112cm\_130737\_p600926 & 196.903 & 60.157 & -14.018 & -14.968 & -16.279 & 0.5853 \\
S251112cm\_121933\_p564412 & 184.886 & 56.737 & -12.457 & -12.829 & -15.146 & 1.3769 \\
S251112cm\_121422\_p560041 & 183.592 & 56.011 & -14.023 & -14.361 & -16.714 & 1.3844 \\
S251112cm\_121215\_p561039 & 183.062 & 56.178 & -13.811 & -14.299 & -16.682 & 1.3935 \\
S251112cm\_120446\_p491112 & 181.192 & 49.187 & -13.193 & -13.603 & -16.119 & 1.4003 \\
S251112cm\_123045\_p571801 & 187.686 & 57.300 & -11.739 & -12.156 & -14.599 & 1.4071 \\
%S251112cm\_123345\_p521517 & 188.437 & 52.255 & -12.487 & -12.940 & -15.440 & 1.4154 \\
%S251112cm\_122243\_p510811 & 185.680 & 51.136 & -13.180 & -13.641 & -15.926 & 1.4221 \\
S251112cm\_121249\_p525412 & 183.204 & 52.903 & -13.338 & -13.798 & -16.142 & 1.4296 \\
%S251112cm\_115035\_p503144 & 177.645 & 50.529 & -12.578 & -13.046 & -15.518 & 1.4371 \\
%S251112cm\_114803\_p461028 & 177.014 & 46.175 & -13.520 & -13.786 & -16.036 & 1.4454 \\
%S251112cm\_114658\_p504208 & 176.740 & 50.702 & -12.569 & -13.049 & -15.313 & 1.4522 \\
%S251112cm\_113448\_p465924 & 173.701 & 46.990 & -13.570 & -13.970 & -16.094 & 1.4596 \\
S251112cm\_111703\_p360829 & 169.264 & 36.141 & -13.224 & -13.670 & -15.558 & 1.4672 \\
%S251112cm\_111451\_p353008 & 168.713 & 35.502 & -12.838 & -13.286 & -15.134 & 1.4747 \\
%S251112cm\_111106\_p433759 & 167.777 & 43.633 & -12.869 & -13.342 & -15.311 & 1.4823 \\
S251112cm\_230658\_m091710 & 346.742 & -9.286 & -14.165 & -14.293 & -16.533 & 1.6392 \\
%S251112cm\_230845\_m123822 & 347.187 & -12.639 & -15.373 & -15.504 & -17.737 & 1.6397 \\
%S251112cm\_225651\_m085803 & 344.212 & -8.968 & -13.162 & -13.359 & -15.535 & 1.6589 \\
S251112cm\_230806\_m055800 & 347.025 & -5.967 & -12.549 & -12.798 & -15.307 & 2.1486 \\
\end{tabular}
\end{table}

\clearpage

\section{DECam Candidates}

Here we summarize all the DECam candidates and their reasons for being rejected as a possible counterpart. Those that could be crossmatched to the legacy survey are listed in Table~\ref{tab:decam_candidates} and those without DR9 hosts are in Table~\ref{tab:decam_candidates_hostless}.

\startlongtable
\begin{deluxetable*}{l|c|c|c|c|c|c}
\tablecaption{A list of identified real DECam discovered transients with hosts crossmatched to LS DR9 \citep{dey_overview_2019}. Their primary reason for being rejected as a possible counterpart to S251112cm is listed. DESI DR1 spec-z's are listed and labeled when available otherwise the median photo-z is given. However, the lower 95th percentile photometric redshift was used to rule out far candidates. \label{tab:decam_candidates}}
\tablehead{
\colhead{DECam ID} & \colhead{TNS Name} & \colhead{R.A. [deg]} & \colhead{Dec [deg]}  & \colhead{$z$} & \colhead{$z_{95}$} & \colhead{Rejection Reason}
}
\startdata
A202511140006389m312701 & 2025admx & 1.662113 & -31.450228 & $0.0360\pm0.0086$ & 0.0211 & AGN-Like \\
A202511140007460m364315 & 2025admy & 1.941743 & -36.720698 & $0.1435\pm0.0263$ & 0.0803 & Far \\
A202511140008036m331239 & 2025admz & 2.015072 & -33.210724 & $0.9017\pm0.3297$ & 0.5195 & Far \\
A202511140011489m335517 & 2025admu & 2.953946 & -33.921294 & $0.3819\pm0.0300$ & 0.3223 & Far \\
A202511140017126m383823 & 2025adjq & 4.302301 & -38.639715 & $1.8491\pm0.4613$ & 0.7886 & Far \\
A202511140022140m380057 & 2025adme & 5.558417 & -38.015839 & $1.0099\pm0.4669$ & 0.0910 & Far \\
A202511140024341m390117 & 2025adpn & 6.142199 & -39.021340 & $0.2486\pm0.0375$ & 0.1929 & Far \\
A202511140039186m423337 & 2025adly & 9.827576 & -42.560403 & $1.3174\pm0.2829$ & 0.5506 & Far \\
A202511140044554m443704 & 2025adlz & 11.230909 & -44.617888 & $0.8860\pm0.1047$ & 0.7530 & Far \\
A202511140048238m425750 & 2025admj & 12.099217 & -42.964003 & $1.3923\pm0.3427$ & 0.6144 & Far \\
A202511140052147m400826 & 2025adjr & 13.061296 & -40.140449 & $0.2532\pm0.0187$ & 0.2247 & Far \\
A202511140056229m414119 & 2025admq & 14.095598 & -41.688607 & $0.9689\pm0.1214$ & 0.7356 & Far \\
A202511140058111m441509 & 2025admh & 14.546315 & -44.252612 & $0.1860\pm0.0240$ & 0.1359 & Far \\
A202511140059559m444449 & 2025adjs & 14.982825 & -44.746866 & $0.2653\pm0.0558$ & 0.2482 & Far \\
A202511140132391m490210 & 2025adjt & 23.163011 & -49.036208 & $0.1363\pm0.0146$ & 0.0983 & Far \\
A202511140210384m515757 & 2025adnd & 32.659813 & -51.965709 & $0.2479\pm0.0344$ & 0.1901 & Far \\
A202511140210496m521700 & 2025adju & 32.706623 & -52.283413 & $0.3760\pm0.2109$ & 0.2793 & Far \\
A202511140211455m542124 & 2025admc & 32.939500 & -54.356637 & $0.3035\pm0.2436$ & 0.1535 & Far \\
A202511140212324m532801 & 2025adjv & 33.135131 & -53.466854 & $1.0264\pm0.0932$ & 0.9003 & Far \\
A202511140214145m540900 & 2025adjw & 33.560500 & -54.150003 & $0.3138\pm0.0281$ & 0.2491 & Far \\
A202511140216255m544127 & 2025adjx & 34.106492 & -54.690914 & $0.1104\pm0.0235$ & 0.0821 & Far \\
A202511142245155m041832 & 2025adna & 341.314382 & -4.308969 & $0.1282\pm0.0378$ & 0.0998 & Far \\
A202511142257452m090047 & 2025admr & 344.438389 & -9.013184 & $0.2544\pm0.0735$ & 0.1887 & Far \\
A202511142301292m112550 & 2025adjy & 345.371760 & -11.430582 & $1.1370\pm0.1869$ & 0.6945 & Far \\
A202511142303272m110203 & 2025admd & 345.863208 & -11.034136 & $0.2812\pm0.3685$ & 0.0163 & SN-Like-Old \\
A202511142304281m100845 & 2025adnb & 346.117073 & -10.145860 & $0.6405\pm0.1044$ & 0.3642 & Far \\
A202511142306290m155754 & 2025adjz & 346.620908 & -15.964994 & $0.2487\pm0.0456$ & 0.1429 & Far \\
A202511142308149m160447 & 2025adnc & 347.062023 & -16.079683 & $0.2478\pm0.0341$ & 0.1756 & Far \\
A202511142319364m112647 & 2025admw & 349.901819 & -11.446323 & $0.2088\pm0.0170$ & 0.1806 & Far \\
A202511142321508m140851 & 2025adne & 350.461597 & -14.147447 & $0.6987\pm0.1050$ & 0.6187 & Far \\
A202511142325202m201213 & 2025adka & 351.334341 & -20.203574 & $0.2334\pm0.3992$ & 0.1285 & Far \\
A202511142338103m293918 & 2025admv & 354.542797 & -29.654888 & $0.1527\pm0.0285$ & 0.0877 & Far \\
A202511142348215m273830 & 2025adkb & 357.089807 & -27.641661 & $0.3238\pm0.0436$ & 0.2352 & Far \\
A202511142350247m273106 & 2025adms & 357.602891 & -27.518351 & $0.2507\pm0.0980$ & 0.1717 & Far \\
A202511142354419m300042 & 2025admt & 358.674735 & -30.011591 & $0.2078\pm0.0472$ & 0.1325 & Far \\
A202511142357140m335527 & 2025adkc & 359.308252 & -33.924256 & $0.8739\pm0.4055$ & 0.7095 & Far \\
A202511150208258m535219 & 2025admk & 32.107507 & -53.872061 & $0.3126\pm0.0696$ & 0.1046 & Far \\
A202511150213021m512012 & 2025admi & 33.258886 & -51.336723 & $0.3332\pm0.0307$ & 0.2722 & Far \\
A202511152250048m003951 & 2025adml & 342.520197 & -0.664072 & $0.3605\pm0.0551$ & 0.2337 & Far \\
C202511140014552m382729 & 2025adqc & 3.730087 & -38.457983 & $0.2472\pm0.1938$ & 0.1263 & Far \\
C202511140017319m330804 & 2025adrj & 4.382867 & -33.134551 & $0.3488\pm0.0594$ & 0.2003 & Far \\
C202511140021364m344727 & 2025adqo & 5.401707 & -34.790796 & $0.2618\pm0.0550$ & 0.1790 & Far \\
C202511140021436m342320 & 2025adkd & 5.431819 & -34.388762 & $0.2469\pm0.0884$ & 0.1269 & Far \\
C202511140024565m345351 & 2025adre & 6.235232 & -34.897409 & $0.3697\pm0.4888$ & 0.0684 & Far \\
C202511140034282m375302 & 2025adke & 8.617397 & -37.883999 & $0.1945\pm0.0485$ & 0.1358 & Far \\
C202511140036387m382613 & 2025adkf & 9.161458 & -38.436880 & $0.1433\pm0.0181$ & 0.1102 & Far \\
C202511140042585m423645 & 2025adlx & 10.743730 & -42.612448 & $1.1352\pm0.2639$ & 0.7004 & Far \\
C202511140044011m425604 & 2025admm & 11.004747 & -42.934397 & $1.6092\pm0.3250$ & 1.0492 & Far \\
C202511140046165m414740 & 2025adoj & 11.568627 & -41.794346 & $0.4820\pm0.2201$ & 0.1720 & Far \\
C202511140046505m422856 & 2025adoz & 11.710347 & -42.482203 & $0.1923\pm0.0272$ & 0.1359 & Far \\
C202511140050048m420726 & 2025adkg & 12.520027 & -42.123906 & $0.1580\pm0.0376$ & 0.1281 & Far \\
C202511140050532m452740 & 2025adqj & 12.721692 & -45.461071 & $0.1058\pm0.0293$ & 0.0585 & Far \\
C202511140052278m443743 & 2025adrf & 13.115880 & -44.628360 & $0.1445\pm0.0793$ & 0.0642 & Far \\
C202511140054061m421033 & 2025adrp & 13.525211 & -42.175914 & $1.0509\pm0.2162$ & 0.6799 & Far \\
C202511140055297m444804 & 2025admo & 13.873949 & -44.801200 & $0.9035\pm0.1870$ & 0.7313 & Far \\
C202511140055320m430041 & 2025admf & 13.883315 & -43.011435 & $0.1405\pm0.0277$ & 0.1089 & Far \\
C202511140055600m441251 & 2025ador & 13.999865 & -44.214132 & $0.1978\pm0.0173$ & 0.1685 & Far \\
C202511140056457m415920 & 2025adir & 14.190461 & -41.988747 & $0.2266\pm0.1039$ & 0.1313 & Far \\
C202511140101108m420417 & 2025adph & 15.295142 & -42.071326 & $0.2817\pm0.1021$ & 0.1132 & Far \\
C202511140105410m440015 & 2025admn & 16.420852 & -44.004210 & $0.0439\pm0.1179$ & 0.0101 & SN-Like-Old \\
C202511140107030m423930 & 2025adnv & 16.762688 & -42.658388 & $0.1411\pm0.0557$ & 0.0452 & Far \\
C202511140109344m430828 & 2025adqy & 17.393174 & -43.141089 & $0.2396\pm0.1061$ & 0.1379 & Far \\
C202511140110393m465519 & 2025admg & 17.663804 & -46.922069 & $0.2630\pm0.0118$ & 0.2466 & Far \\
C202511140118195m472436 & 2025admp & 19.581304 & -47.410027 & $0.4855\pm0.0561$ & 0.3800 & Far \\
C202511140127131m465750 & 2025adpo & 21.804496 & -46.963889 & $0.2665\pm0.0344$ & 0.2129 & Far \\
C202511140211093m513710 & 2025adnq & 32.788549 & -51.619344 & $0.2086\pm0.0461$ & 0.1398 & Far \\
C202511140211306m513945 & 2025adpr & 32.877351 & -51.662567 & $0.4521\pm0.0777$ & 0.2825 & Far \\
C202511140221449m550650 & 2025adkh & 35.437032 & -55.113909 & $0.6929\pm0.0755$ & 0.5435 & Far \\
C202511142248197m051604 & 2025adqq & 342.082190 & -5.267853 & $0.1528\pm0.0303$ & 0.1057 & Far \\
C202511142257439m130705 & 2025adqb & 344.432748 & -13.118006 & $0.2896\pm0.1669$ & 0.0640 & Far \\
C202511142302526m132518 & 2025adrg & 345.719030 & -13.421754 & $0.1556\pm0.0126$ & 0.1408 & Far \\
C202511142304092m082845 & 2025adoc & 346.038187 & -8.479193 & $0.1744\pm0.0093$ & 0.1592 & Far \\
C202511142344126m304316 & 2025adpi & 356.052420 & -30.721090 & -- & -- & AGN-Like \\
C202511142350343m300604 & 2025adps & 357.642953 & -30.101076 & $0.1592\pm0.0495$ & 0.0439 & Far \\
C202511142350481m304211 & 2025adok & 357.700395 & -30.703034 & $0.2781\pm0.1619$ & 0.1450 & Far \\
C202511160043490m431706 & 2025adsk & 10.954355 & -43.284995 & $0.6999\pm0.0710$ & 0.5124 & Far \\
C202511160044250m430417 & 2025adhe & 11.103609 & -43.071046 & $0.1216\pm0.0052$ & 0.1116 & Far \\
T202511140003508m331219 & 2025adpt & 0.961726 & -33.205275 & $0.2680\pm0.0251$ & 0.2248 & Far \\
T202511140004157m342211 & 2025adod & 1.065351 & -34.369725 & $0.5949\pm0.0940$ & 0.3830 & Far \\
T202511140010116m332613 & 2025adpu & 2.548331 & -33.436966 & $0.1197\pm0.0405$ & 0.0583 & Far \\
T202511140011395m323754 & 2025adqr & 2.914781 & -32.631587 & $0.2733_{\text{spec}}$ & 0.2644 & far \\
T202511140013387m393329 & 2025adki & 3.411403 & -39.558040 & $0.3009\pm0.4219$ & 0.1141 & Far \\
T202511140014036m373615 & 2025adrk & 3.514959 & -37.604054 & $0.0530\pm0.0136$ & 0.0279 & SN-Like-Old \\
T202511140015272m402106 & 2025adkj & 3.863456 & -40.351642 & $0.0421\pm0.0095$ & $0.092_{\text{spec}}$ & far \\
T202511140016373m343813 & 2025adkk & 4.155565 & -34.636995 & $0.0768\pm0.2497$ & 0.0154 & SN-Like-Old \\
%T202511140018095m335115 & 2025adjp & 4.539708 & -33.854242 & $0.0393\pm0.0363$ & 0.0111 &  \\
T202511140018462m334111 & 2025adqd & 4.692482 & -33.686424 & $0.9668\pm0.1244$ & 0.7538 & Far \\
T202511140020414m392556 & 2025adol & 5.172411 & -39.432321 & $0.0578\pm0.0240$ & 0.0223 & SN-Like-Old \\
T202511140020450m380123 & 2025adnf & 5.187579 & -38.023064 & $0.3298\pm0.1153$ & 0.0559 & Far \\
T202511140023013m370552 & 2025adns & 5.755332 & -37.097767 & $0.3083\pm0.4663$ & 0.0817 & Far \\
T202511140028381m361629 & 2025adhg & 7.158625 & -36.274600 & $0.1195\pm0.0577$ & 0.0634 & Far \\
T202511140029283m414027 & 2025adnl & 7.367930 & -41.674292 & $1.2246\pm0.4063$ & 0.1950 & Far \\
T202511140030086m410443 & 2025adkl & 7.535766 & -41.078709 & $0.1457\pm0.0325$ & 0.0875 & Far \\
T202511140030555m393903 & 2025adkm & 7.731292 & -39.650733 & $0.0891\pm0.0113$ & 0.0648 & Far \\
T202511140030556m414018 & 2025adqz & 7.731546 & -41.671734 & $0.5838\pm0.0702$ & 0.4485 & Far \\
T202511140031492m410519 & 2025adpb & 7.954879 & -41.088569 & $0.4484\pm0.0688$ & 0.3390 & Far \\
T202511140034296m413931 & 2025adkn & 8.623334 & -41.658603 & $0.4957\pm0.2419$ & 0.2313 & Far \\
T202511140042227m415428 & 2025adko & 10.594384 & -41.907774 & $0.8857\pm0.1601$ & 0.6521 & Far \\
T202511140043457m425620 & 2025adpa & 10.940529 & -42.938834 & $0.5973\pm0.2495$ & 0.1307 & Far \\
T202511140047331m405028 & 2025adkp & 11.887936 & -40.841148 & $0.1610\pm0.0139$ & 0.1287 & Far \\
T202511140049053m405309 & 2025adon & 12.271927 & -40.885832 & $0.4761\pm0.1545$ & 0.2921 & Far \\
T202511140052584m435834 & 2025adnt & 13.243201 & -43.975979 & $0.7984\pm0.0543$ & 0.7079 & Far \\
T202511140053580m445337 & 2025adpy & 13.491477 & -44.893640 & $0.3197\pm0.5133$ & 0.0449 & Far \\
T202511140054444m463660 & 2025adpz & 13.685130 & -46.616634 & $0.7448\pm0.1074$ & 0.4921 & Far \\
T202511140055191m403254 & 2025ados & 13.829776 & -40.548425 & $0.2489\pm0.0327$ & 0.1909 & Far \\
T202511140055346m463810 & 2025adkq & 13.894152 & -46.636135 & $0.4054\pm0.1662$ & 0.1999 & Far \\
T202511140057592m453316 & 2025adpp & 14.496842 & -45.554414 & $0.0521\pm0.0174$ & 0.0302 & SN-Like-Old \\
T202511140100184m464928 & 2025adqk & 15.076744 & -46.824499 & $0.8710\pm0.0386$ & 0.7967 & Far \\
T202511140103535m462507 & 2025adnp & 15.973059 & -46.418720 & $0.0870\pm0.0259$ & 0.0528 & Far \\
T202511140111494m451919 & 2025adkr & 17.955927 & -45.321882 & $0.0910\pm0.0087$ & 0.0736 & Far \\
T202511140113198m470350 & 2025adng & 18.332417 & -47.063929 & $0.9485\pm0.1696$ & 0.7008 & Far \\
T202511140115136m474104 & 2025adks & 18.806510 & -47.684444 & $0.0732\pm0.0253$ & 0.0376 & SN-Like-Old \\
T202511140116542m471802 & 2025adql & 19.225832 & -47.300574 & $0.1249\pm0.0651$ & 0.0798 & Far \\
T202511140123546m480240 & 2025adkt & 20.977666 & -48.044455 & $0.1295\pm0.0163$ & 0.0986 & Far \\
T202511140127319m462509 & 2025adqm & 21.882824 & -46.419196 & $0.1340\pm0.0636$ & 0.0868 & Far \\
T202511140209066m514909 & 2025adnh & 32.277607 & -51.819067 & $1.1099\pm0.2911$ & 0.4065 & Far \\
T202511140210579m512140 & 2025adku & 32.741218 & -51.361041 & $0.1547\pm0.0230$ & 0.1072 & Far \\
T202511140214207m542258 & 2025adkv & 33.586181 & -54.382904 & $0.2988\pm0.0215$ & 0.2601 & Far \\
T202511140214229m535960 & 2025adqp & 33.595250 & -53.999964 & $0.6444\pm0.2141$ & 0.3461 & Far \\
T202511140229395m501154 & 2025adkw & 37.414749 & -50.198432 & $0.5433\pm0.1494$ & 0.3126 & Far \\
T202511140231254m530740 & 2025adni & 37.855953 & -53.127895 & $0.3157\pm0.0263$ & 0.2713 & Far \\
T202511142245384m031402 & 2025adkx & 341.410020 & -3.233758 & $0.2165\pm0.0288$ & 0.1555 & Far \\
T202511142248554m003651 & 2025adqe & 342.230750 & -0.614167 & $0.2601_{\text{spec}}$ & 0.2590 & far \\
T202511142252028m071541 & 2025adjf & 343.011598 & -7.261509 & $0.1325\pm0.0126$ & 0.1120 & Far \\
T202511142259296m082951 & 2025adot & 344.873449 & -8.497500 & $0.6298\pm0.0397$ & 0.5630 & Far \\
T202511142259331m092758 & 2025adqf & 344.888125 & -9.466225 & $0.2188\pm0.0184$ & 0.1776 & Far \\
T202511142300078m085830 & 2025adnw & 345.032456 & -8.974967 & $0.1076\pm0.3411$ & 0.0074 & No Evolution \\
T202511142301363m120722 & 2025adnu & 345.401119 & -12.122680 & $0.4982\pm0.0282$ & 0.4189 & Far \\
T202511142306403m081116 & 2025adky & 346.668151 & -8.187785 & $0.2476\pm0.1226$ & 0.1006 & Far \\
T202511142312514m121611 & 2025adro & 348.214261 & -12.269704 & $0.3169\pm0.2915$ & 0.0108 & SN-Like-Old \\
T202511142313237m102328 & 2025adra & 348.348902 & -10.391036 & $0.0570\pm0.0199$ & 0.0259 & SN-Like-Old \\
T202511142314384m093453 & 2025adrh & 348.660023 & -9.581275 & $0.4901\pm0.1055$ & 0.2918 & Far \\
T202511142322507m145456 & 2025adny & 350.711386 & -14.915487 & $0.3087\pm0.0351$ & 0.2129 & Far \\
T202511142323401m144758 & 2025adnr & 350.917103 & -14.799449 & $0.1244\pm0.0504$ & 0.0572 & Far \\
T202511142323419m195742 & 2025adop & 350.924797 & -19.961606 & $0.1805\pm0.0539$ & 0.0935 & Far \\
T202511142324462m215535 & 2025adqa & 351.192467 & -21.926292 & $0.0496\pm0.0616$ & 0.0066 & SN-Like-Old \\
T202511142325566m203923 & 2025adpf & 351.485696 & -20.656463 & $0.5554\pm0.6567$ & 0.0236 & No Evolution \\
T202511142327265m195030 & 2025adrc & 351.860481 & -19.841717 & $0.3979\pm0.1058$ & 0.2399 & Far \\
T202511142329282m211934 & 2025adrm & 352.367413 & -21.326026 & $0.7312\pm0.4857$ & 0.3305 & Far \\
T202511142344510m294635 & 2025adrn & 356.212349 & -29.776400 & $0.1629\pm0.0315$ & 0.1096 & Far \\
T202511142348239m295503 & 2025adoy & 357.099548 & -29.917400 & $0.9600\pm0.1670$ & 0.6385 & Far \\
T202511142348448m324601 & 2025adqv & 357.186556 & -32.766928 & -- & -- & AGN-Like \\
T202511142353140m335851 & 2025adpm & 358.308527 & -33.980818 & $0.0857\pm0.0406$ & 0.0264 & No Evolution \\
T202511142354127m314607 & 2025adoi & 358.552805 & -31.768573 & -- & -- & AGN-Like \\
T202511142355571m283648 & 2025adqw & 358.988105 & -28.613414 & $0.2967\pm0.0220$ & 0.2715 & Far \\
T202511142356038m283558 & 2025adqi & 359.016032 & -28.599309 & $0.2967\pm0.0220$ & 0.2715 & Far \\
T202511142356277m332052 & 2025adpq & 359.115427 & -33.347745 & $0.1925\pm0.0216$ & 0.1564 & Far \\
T202511142358255m303845 & 2025adiq & 359.606365 & -30.645888 & $0.0874\pm0.0257$ & 0.0442 & Far \\
T202511142358392m334230 & 2025adqx & 359.663310 & -33.708289 & $0.5866\pm0.4739$ & 0.1245 & Far \\
T202511142358418m335938 & 2025adlh & 359.674348 & -33.994014 & $0.5183\pm0.0321$ & 0.4455 & Far \\
T202511160043404m440927 & 2025adsl & 10.918533 & -44.157374 & $0.6792\pm0.0782$ & 0.5276 & Far \\
\enddata
\end{deluxetable*}

\begin{deluxetable*}{l|l|l|l|l}
\tablecaption{A list of identified real DECam discovered transients without hosts crossmatched to LS DR9 \citep{dey_overview_2019}. Photometric redshifts are derived from \citet{beck_photometric_2016}, \citet{beck_wise-ps1-strm_2022}, and PS-z \label{tab:decam_candidates_hostless}}
\tablehead{
\colhead{DECam ID} & \colhead{TNS Name} & \colhead{R.A. [deg]} & \colhead{Dec [deg]} & \colhead{Rejection Reason}
}
\startdata
T202511142320394m194332 & 2025adlb & 350.164162 & -19.725628 & AGN-Like \\
T202511142336538m264738 & 2025adpd & 354.224152 & -26.793910 & AGN-Like \\
T202511142328007m181603 & 2025adoo & 352.002884 & -18.267539 & AGN-Like \\
T202511142323040m195152 & 2025adrb & 350.766682 & -19.864251 & SN-Like-Old \\
T202511142314555m185016 & 2025adrl & 348.731253 & -18.837769 & Far $(z_\text{phot} = 0.244)$\\
T202511142325219m154649 & 2025adno & 351.341323 & -15.780296 & SN-Like-Old \\
T202511142316181m195535 & 2025adkz & 349.075524 & -19.926385 & Far $(z_\text{phot} = 0.200)$\\
%GOTO25jzt, T202511142338118m254749, ATLAS25ogu, PS25mtr & 2025adiz & 354.549229 & -25.796956 &  \\
T202511142327143m180558 & 2025adri & 351.809377 & -18.099573 & Far $(z_\text{phot} = 0.425)$ \\
T202511142333552m234340 & 2025adrd & 353.480142 & -23.727818 & Far $(z_\text{phot} = 0.0473)$ \\
T202511142321491m170031 & 2025adqu & 350.454380 & -17.008569 & AGN-Like \\
T202511142318546m191013 & 2025adqt & 349.727695 & -19.170170 & No Evolution \\
T202511142318085m194819 & 2025adqs & 349.535310 & -19.805392 & Far $(z_\text{phot} = 0.210)$ \\
T202511142340113m271353 & 2025adqn & 355.047172 & -27.231332 & No Evolution \\
T202511142346409m250413 & 2025adqh & 356.670354 & -25.070385 & No Evolution \\
T202511142316052m202752 & 2025adqg & 349.021512 & -20.464335 & AGN-Like \\
T202511142347345m260925 & 2025adpx & 356.893950 & -26.157063 & No Evolution \\
T202511142329207m260829 & 2025adpw & 352.336336 & -26.141425 & No Evolution \\
T202511142328006m241756 & 2025adpv & 352.002428 & -24.298937 & No Evolution \\
T202511142337415m274307 & 2025adpl & 354.423068 & -27.718672 & Far $(z_\text{phot} = 0.328)$ \\
T202511142332043m201604 & 2025adpk & 353.017808 & -20.267773 & Far $(z_\text{phot} = 0.205)$ \\
T202511142316135m203232 & 2025adpj & 349.056065 & -20.542170 & Far $(z_\text{phot} = 0.079)$ \\
T202511142339473m251308 & 2025adpg & 354.947017 & -25.218999 & No Evolution \\
T202511142310550m164348 & 2025adpe & 347.729167 & -16.730020 & AGN-Like \\
T202511142310393m170215 & 2025adpc & 347.663574 & -17.037568 & No Evolution \\
T202511142328218m235112 & 2025adox & 352.091033 & -23.853462 & Far $(z_\text{phot} = 0.123)$ \\
T202511142325245m160906 & 2025adow & 351.352105 & -16.151802 & Far $(z_\text{phot} = 0.630)$ \\
T202511142323592m235840 & 2025adov & 350.996595 & -23.977743 & Far $(z_\text{phot} = 0.691)$ \\
T202511142323279m232146 & 2025adou & 350.866330 & -23.362880 & No Evolution \\
T202511142323279m232508 & 2025adoq & 350.866056 & -23.418823 & Far $(z_\text{phot} = 0.089)$ \\
T202511142333035m221733 & 2025adom & 353.264737 & -22.292616 & Far $(z_\text{phot} = 0.230)$ \\
T202511142350310m260648 & 2025adoh & 357.629370 & -26.113296 & No Evolution \\
T202511142328473m172010 & 2025adog & 352.197119 & -17.336075 & Far $(z_\text{phot} = 0.121)$ \\
T202511142328467m194413 & 2025adof & 352.194505 & -19.737005 & Far $(z_\text{phot} = 0.143)$ \\
T202511142328011m160955 & 2025adoe & 352.004588 & -16.165380 & Far $(z_\text{phot} = 0.628)$  \\
T202511142326455m232708 & 2025adob & 351.689641 & -23.452259 & Far $(z_\text{phot} = 0.057)$ \\
T202511142331314m221524 & 2025adoa & 352.880780 & -22.256663 & No Evolution \\
T202511142323288m185854 & 2025adnz & 350.870126 & -18.981659 & Far $(z_\text{phot} = 0.144)$ \\
T202511142312568m184645 & 2025adnx & 348.236788 & -18.779247 & Slow Decline \\
T202511142319148m163240 & 2025adnn & 349.811531 & -16.544583 & Far $(z_\text{phot} = 0.167)$ \\
T202511142315273m194918 & 2025adnm & 348.863910 & -19.821728 & No Evolution \\
T202511142322179m220034 & 2025adnk & 350.574446 & -22.009574 & AGN-Like \\
T202511142310143m171234 & 2025adnj & 347.559791 & -17.209528 & No Evolution \\
T202511142317242m190555 & 2025adlg & 349.350857 & -19.098733 & Far $(z_\text{phot} = 0.195)$ \\
T202511142340091m242417 & 2025adlf & 355.038030 & -24.404643 & AGN-Like \\
T202511142336089m221231 & 2025adle & 354.036939 & -22.208691 & AGN-Like \\
T202511142333154m221641 & 2025adld & 353.313992 & -22.278125 & AGN-Like \\
T202511142323533m180247 & 2025adlc & 350.972226 & -18.046371 & No Evolution \\
T202511142316208m164448 & 2025adla & 349.086577 & -16.746547 & AGN-Like \\
\enddata
\end{deluxetable*}

\section{ZTF Candidates}

Here in table~\ref{tab:ztf_candidates} we present a table of all the candidates identified by the Zwickty Transient Facility (ZTF) that pass the cuts as outlined in Section~\ref{sec:candidates}.

\startlongtable
\begin{deluxetable*}{l|c|c|c|c}
\tablecaption{Here is a table of all the candidates observed by the Zwicky Transient Facility. *This SN Ia was classified with a photometric fit to various SN models. We find that the SN Ia model fits with a $\chi^2$/d.o.f. that is at least 4.8 times better than any other model. \label{tab:ztf_candidates}}
\tablehead{
\colhead{ZTF ID} & \colhead{TNS Name} & \colhead{R.A. [deg]} & \colhead{Dec [deg]} & \colhead{Rejection Reason}
}
\startdata
ZTF25acdywhh & \nodata & 163.785429 & 27.731521 & Bogus \\
ZTF25acebput & 2025adhy & 339.583771 & -0.954412 & Far (Photo-z $= 0.138\pm0.017$) \\
ZTF25aceeinw & \nodata & 170.837117 & 37.822862 & Bogus \\
ZTF25aceejue & \nodata & 173.598218 & 43.223559 & Far (Photo-z $= 0.199\pm0.016$) \\
ZTF25aceekrn & 2025adht & 180.101460 & 49.047764 & Far (Photo-z $= 0.139\pm0.033$) \\
ZTF25aceekrw & 2025adxs & 180.737425 & 48.352337 & Far (Photo-z $= 0.127\pm0.044$) \\
ZTF25aceeksk & \nodata & 172.553987 & 49.158452 & Far (Photo-z $= 0.668\pm0.344$) \\
ZTF25aceekwp & 2025adhu & 181.804049 & 55.653636 & Far (Photo-z $= 0.591\pm0.235$) \\
ZTF25aceekzz & 2025adib & 169.076784 & 39.863378 & AGN \\
ZTF25aceelbm & 2025adic & 169.544424 & 41.094706 & Far (Photo-z $= 0.210\pm0.056$) \\
ZTF25aceelca & \nodata & 168.407785 & 38.899913 & Far (Photo-z $= 1.054\pm0.142$) \\
ZTF25aceelgz & 2025adid & 190.811191 & 57.328899 & AGN \\
ZTF25aceelhe & \nodata & 194.192011 & 52.569014 & AGN \\
ZTF25aceellw & \nodata & 192.195160 & 63.116948 & AGN \\
ZTF25aceelmh & 2025adhz & 184.916369 & 57.167778 & AGN \\
ZTF25aceelmm & \nodata & 196.133708 & 60.724683 & AGN \\
ZTF25aceelrj & \nodata & 199.705220 & 59.783998 & AGN \\
ZTF25aceelxx & \nodata & 176.605392 & 39.114657 & Declining pre-detections\\ %Far (Photo-z $= 0.072\pm0.0237$) \\
ZTF25aceemaz & \nodata & 178.227673 & 40.921501 & AGN \\
ZTF25aceembk & \nodata & 197.218544 & 62.610051 & No Evolution \\
ZTF25aceemie & \nodata & 196.190000 & 53.161038 & AGN \\
ZTF25aceemla & 2025adie & 185.395966 & 56.586273 & Far (Photo-z $= 0.665\pm0.239$) \\
ZTF25aceempf & \nodata & 180.138894 & 48.493581 & Far (Photo-z $= 0.084\pm0.021$) \\
ZTF25aceempu & 2025adia & 169.569004 & 35.985827 & Far (Photo-z $= 0.410\pm0.099$) \\
ZTF25aceemre & \nodata & 175.915968 & 34.751993 & Far (Photo-z $= 0.829\pm0.069$) \\
%ZTF25aceeneu & 2025adtq & 175.614191 & 34.008569 &  \\
ZTF25aceevwg & 2025adln & 168.037357 & 30.426614 & SN Ia* \\
\enddata

\end{deluxetable*}

\section{Spectroscopic Follow-up}

Here in Table~\ref{tab:salt_spec}, we summarize the results of the spectroscopic follow-up of transients with the South African Large Telescope (SALT). SN 2025adpq is a SN Ia in a collisional ring and was presented in \citet{oconnor_sn_2026}.

\begin{table*}
\centering
\caption{A table of transients that were followed-up with the SALT GW program.\label{tab:salt_spec}}
\begin{tabular}{c|c|c|c|c|c}
\hline
Name & Internal Name & Classification & $\text{P}_\text{2D}$ & $z$ & Rejection Reason \\
\hline
2025adcv & -- & SN Ia & 0.608 & 0.089 & Far \\
2025adcy & -- & SN Ia & 0.931 & 0.146 & Far \\
2025addc & -- & SN Ia & 0.966 & 0.145 & Far \\
2025adgp & -- & SN Ia & 0.666 & 0.048 & Far \\
2025adjp & T202511140018095m335115 & Galaxy & 0.416 & -- & Far (Photo-z) \\
2025adkl & T202511140030086m410443 & Galaxy & 0.625 & -- & Far (Photo-z) \\
2025adpq & T202511142356277m332052 & SN Ia* & 0.706 & 0.1540 & Far \\
2025adin & -- & Galaxy & 0.301 & 0.159 & Far \\
2025aimh & -- & Galaxy & 0.636 & 0.279 & Far \\
%2025yvp  & -- &  &  &  &  & \\

\end{tabular}
\end{table*}

\section{Classified Candidates on TNS}

Here we present the publicly classified candidates from the TNS within the 2D localization area from the LIGO event (Table~\ref{tab:tns_candidates}). 

\begin{table*}
\centering
\caption{A list of all classified transients (not including SN 2025adtq) or transients with reported redshift from TNS within the $99\%$ probability region and with a discovery date later than 2025-11-08 15:18:45 UTC (T$-4$ days) and before 2025-11-26 (T$+14$ days). ``Too Young" incidates that the LIGO event happened before the approximated $t_0$ of the SN, and it therefore is not a viable SKN candidate. \label{tab:tns_candidates}}
\begin{tabular}{c|c|c|c|c|c|c}
\hline
Name & Internal Name & Classification & $\text{P}_\text{2D}$ & $z$ & $d_L$ (Mpc) & Rejection Reason \\
\hline
%2025accy & -- & SN Ia & 0.936 & 0.0441 & 186.9 & Far\\
%2025acsj & -- & SN Ia & 0.735 & 0.0500 & 213.0 & Far\\
%2025acvj & -- & SN Ia   & 0.807 & 0.0304 & 127.8  & Ia \\
%2025adcv & -- & SN Ia   & 0.608 & 0.0890 & 389.7 & Far \\
%2025adcy & -- & SN Ia   & 0.931 & 0.1460 & 663.6 & Far \\
%2025addc & -- & SN Ia   & 0.966 & 0.1450 & 658.6 & Far \\
%2025adgp & -- & SN Ia   & 0.666 & 0.0478 & 203.3  & Far \\
2025adgq & -- & SN Ib   & 0.280 & 0.0643 & 276.8  & Far \\
2025adhf & -- & SN Ia   & 0.746 & 0.0980 & 431.7  & Far \\
2025adhs & -- & SLSN-II & 0.968 & 0.0427 & 180.8  & Far \\
2025adim & -- & SN Ia   & 0.685 & 0.0910 & 399.0  & Far \\
2025adiw & -- & SN Ia   & 0.311 & 0.1650 & 758.9  & Far \\
2025adiz & T202511142338118m254749 & SN Ia   & 0.391 & 0.1300 & 584.9  & Far \\
2025adjc & -- & SN Ia   & 0.262 & 0.0286 & 119.8  & Ia \\
%2025adpq & T202511142356277m332052 & None    & 0.706 & 0.1540 & 703.5  & Far \\
2025adrr & -- & SN Ia   & 0.786 & 0.0560 & 239.6  & Far \\
%2025adtq & -- & SN IIb  & 0.737 & 0.0333 & 140.2  & This Work \\
2025aedy & -- & SN IIb  & 0.975 & 0.0280 & 117.4  & Too Young \\
2025aegk & -- & SN Ib   & 0.984 & 0.0184 & 76.6   & Too Young \\
2025aept & ZTF25aceglpp & SN II   & 0.546 & 0.0307 & 129.1  & Normal II \\
2025afad & ZTF25accmopr & SLSN-I  & 0.820 & 0.3070 & 1528.3 & Far \\
2025afhg & ZTF25acffrxl & SN Ic   & 0.063 & 0.0203 & 84.8 & Too Young \\
2025altn & -- & None    & 0.947 & 1.0560 & 6774.1 & Far \\
\end{tabular}
\end{table*}

\section{Comparison to Existing Populations}
\label{ap:iib_pop}

\begin{figure*}
    \centering
    \includegraphics[width=0.98\linewidth]{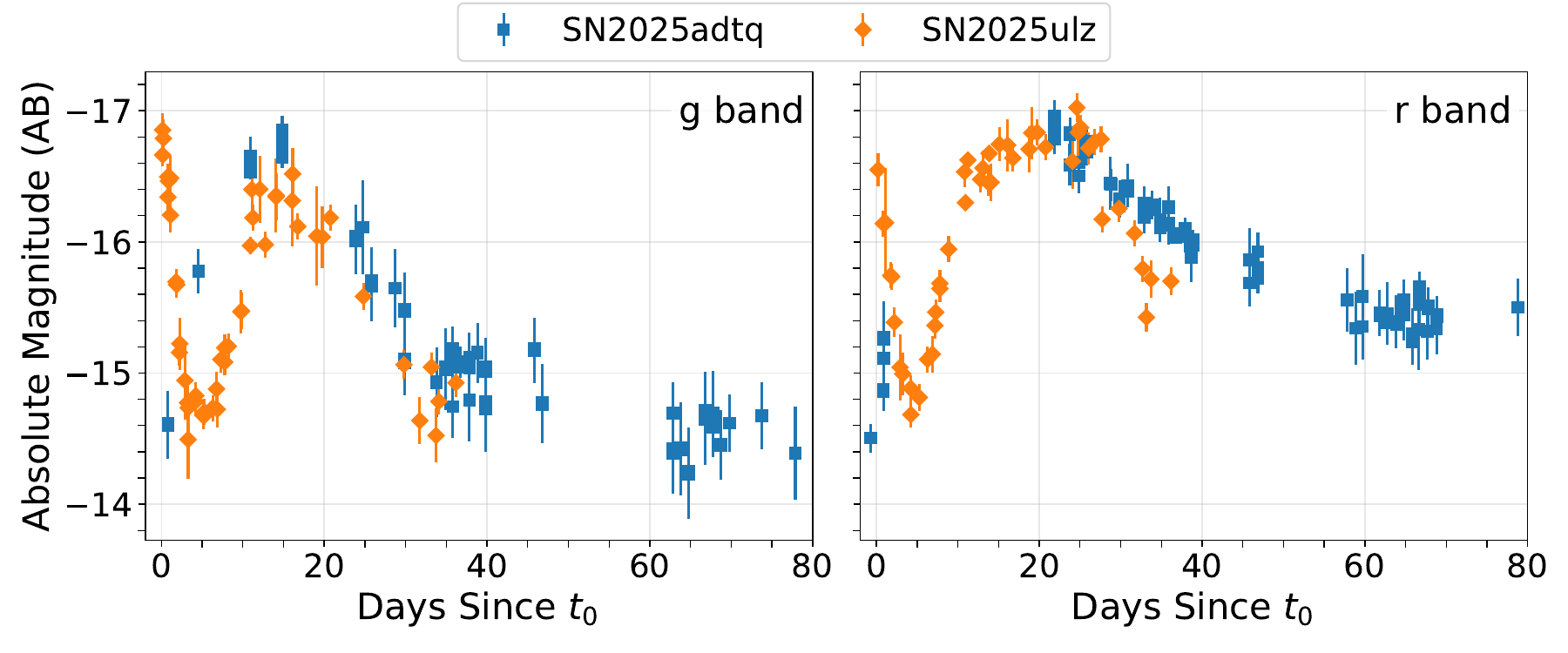}
    \caption{Here we compare the $g$ and $r$ band observations from SN 2025ulz and SN 2025adtq.}
    \label{fig:LC_comparison}
\end{figure*}

\begin{figure*}
    \centering
    \includegraphics[width=0.7\linewidth]{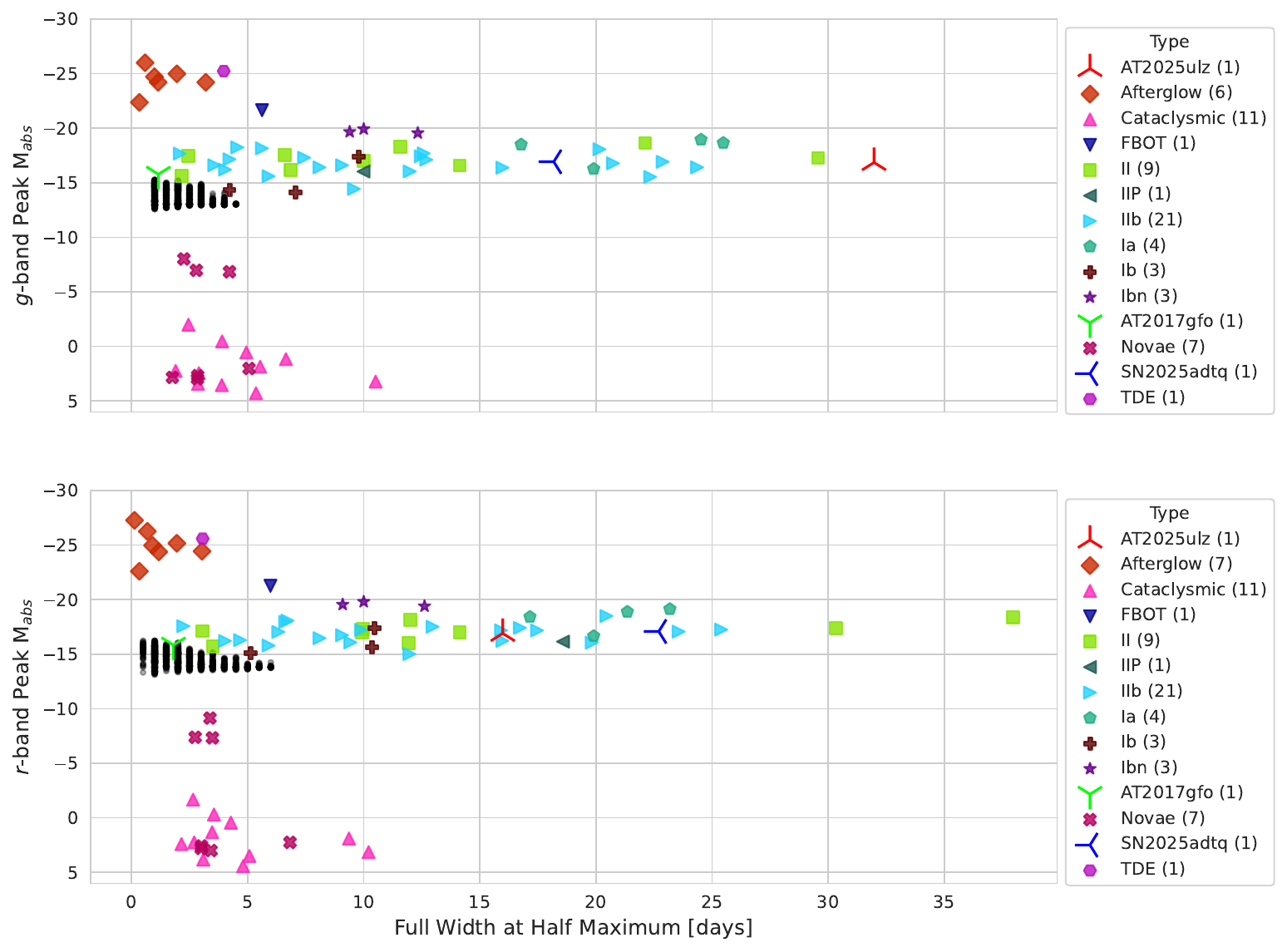}
    \includegraphics[width=0.7\linewidth]{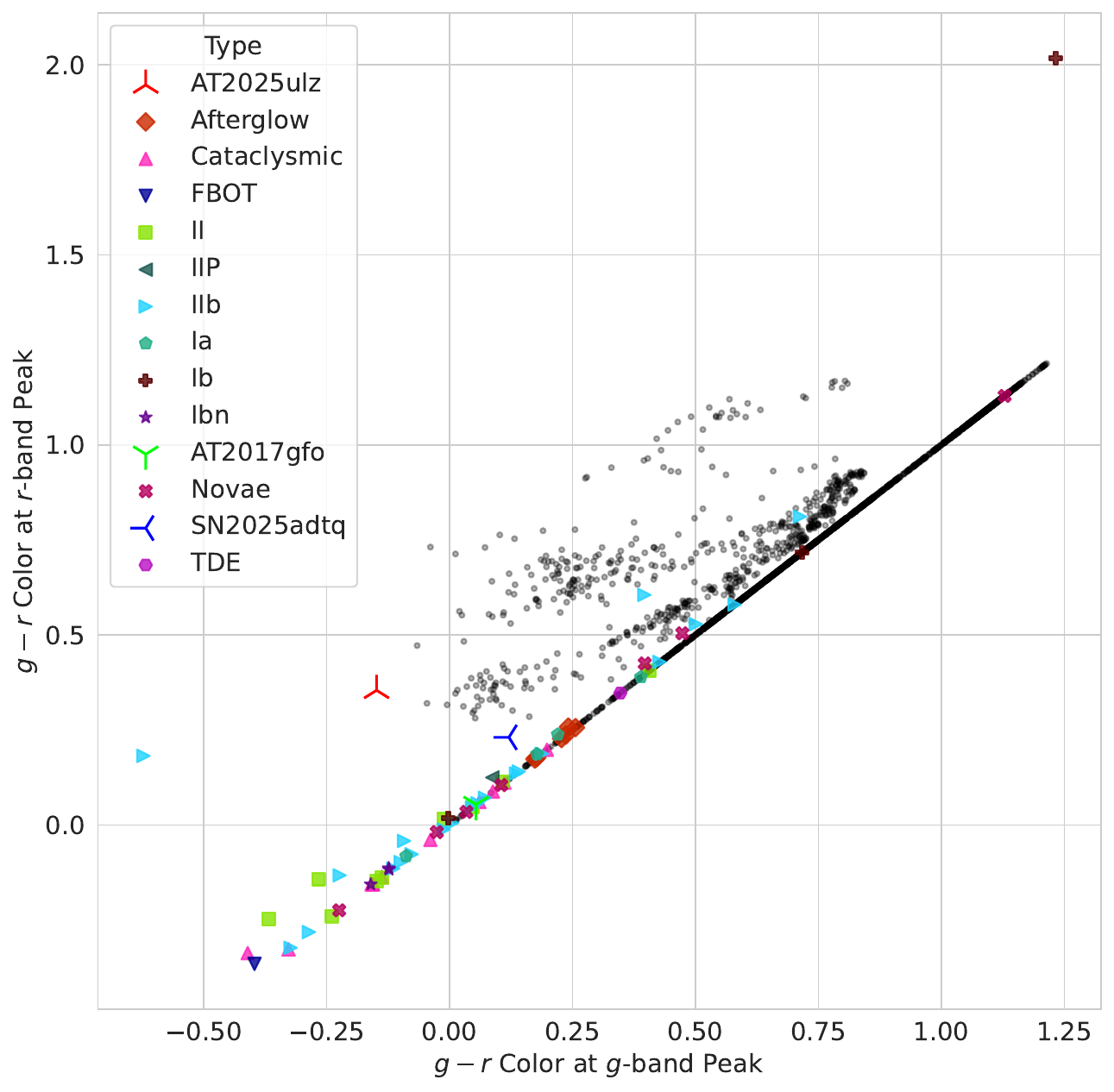}
    \caption{Full Width Half Maximum (FWHM) of a population of IIbs versus the $g$ (top) and $r$-band (middle) peak. Bottom: $g-r$ color at $g$-band peak versus the $g-r$ color at the $r$-band peak. The black dots represent a grid of modeled KNe. The method of analysis and background populations are from \citet{barna_iib_2025}}.
    \label{fig:barna_analsyis}
\end{figure*}

We compare the lightcurves of SN 2025ulz and SN 2025adtq in figure~\ref{fig:LC_comparison}. Then in figure~\ref{fig:barna_analsyis} we compare the SN 2025adtq to SN 2025ulz and to the greater population of IIb SN and model KN from \citet{barna_iib_2025}.

\bibliography{sample701}{}
\bibliographystyle{aasjournalv7}

%% This command is needed to show the entire author+affiliation list when
%% the collaboration and author truncation commands are used.  It has to
%% go at the end of the manuscript.
%\allauthors

%% Include this line if you are using the \added, \replaced, \deleted
%% commands to see a summary list of all changes at the end of the article.
%\listofchanges

\end{document}